\documentclass[aps,twocolumn]{revtex4}
\pdfoutput=1

\usepackage{eurosym}
\usepackage{amsfonts}
\usepackage{amsmath}
\usepackage{amssymb,epsf}
\usepackage{color}
\usepackage{epstopdf}
\usepackage{graphicx}
\usepackage{caption}
\usepackage{subfig}
\usepackage{hyperref}
\usepackage{mathtools}
\usepackage{lipsum}
\usepackage{floatrow}
\definecolor{aogreen}{rgb}{0.0, 0.5, 0.0}

\def\ketm#1{  \left\vert  #1   \right\rangle   }

\def\bram#1{  \left\langle  #1   \right\vert   }

\begin{document}

\title{Controllable simulation of topological phases and edge states with quantum walk}

\author{S. Panahiyan $^{1,2,3}$}
\email{email address: shahram.panahiyan@uni-jena.de}
\author{S. Fritzsche $^{1,2,3}$ }
\email{email address: s.fritzsche@gsi.de}
\affiliation{$^1$Helmholtz-Institut Jena, Fr\"{o}belstieg 3, D-07743 Jena, Germany  \\
	$^2$GSI Helmholtzzentrum f\"{u}r Schwerionenforschung, D-64291 Darmstadt, Germany \\
	$^3$Theoretisch-Physikalisches Institut, Friedrich-Schiller-University Jena, D-07743 Jena, Germany}

\date{\today}

\begin{abstract}
We simulate various topological phenomena in condense matter, such as formation of different topological phases, boundary and edge states, through two types of quantum walk with step-dependent coins. Particularly, we show that one-dimensional quantum walk with step-dependent coin simulates all types of topological phases in BDI family, as well as all types of boundary and edge states. In addition, we show that step-dependent coins provide the number of steps as a controlling factor over the simulations. In fact, with tuning number of steps, we can determine the occurrences of boundary, edge states and topological phases, their types and where they should be located. These two features make quantum walks versatile and highly controllable simulators of topological phases, boundary, edge states, and topological phase transitions. We also report on emergences of cell-like structures for simulated topological phenomena. Each cell contains all types of boundary (edge) states and topological phases of BDI family.
\end{abstract}

\maketitle
\section{Introduction} \label{Intro}

Phenomena, such as the integer Hall effect \cite{Thouless}, fractional charges \cite{Qi2008} and topological insulators \cite{Hasan,Qi,Fidkowski,Koenig}, emerge in different phases of matter known as topological phases in the literature. In contrast to conventional phases of matter, the topological phases are symmetry-preserving and can be parametrized by global topological orders (topological invariants). These topological invariants characterize the global structure of ground-state wave function of matters and do not change continuously \cite{Thouless}. The topological phases are separated by boundary states where topological invariant is usually ill-defined and the gapped energy bands close their gap \cite{Thouless}. In general, there are three types of gap-closer for the energy bands (three types of boundary and edge states) \cite{Fidkowski,Schnyder,Chiu,Jin,Zhao2012}: a) linearly closing (linear dispersive behavior) which is known as Dirac cone, b) nonlinearly closing (nonlinear dispersive behavior) which is called Fermi arc and c) flat closing (dispersionless behavior) which is named flat band.

So far, topological phases and boundary states have been reported in topological insulators and superconductors \cite{Hasan,Qi,Fidkowski,Koenig,Hsieh}, cold atoms in optical lattices \cite{Atala,Leder}, and phononic states in mechanical oscillators \cite{Susstrunk}. Recently, it has been reported that quantum walks with photons can be used as simulators of the topological phenomena \cite{KitagawaExp,Cardano,Wang2018,Cardano2017,Barkhofen,Flurin,Zhan,Xiao,Nitsche,Wang}. 

Quantum walks are universal computational primitives \cite{Lovett} that have been applied to simulate different quantum systems and phenomena \cite{Mohseni,Vakulchyk} including topological phases as observed in condensed matter \cite{Kitagawa}. In particular, quantum walks can simulate all types of topological phases in one and two dimensions \cite{Kitagawa,Kitagawa2012,Asboth,Obuse,Chen}. In addition, quantum walks enable one to extract topological invariants \cite{Ramasesh}, create anomalous Floquet Chern insulators \cite{Sajid}, investigate bulk-boundary correspondence \cite{Asboth2013} and topological phase transitions \cite{Rakovszky,Mera,Panahiyan2019}. The power of quantum walks as simulators arises from their flexibility and controllability. In fact, quantum walks are able to check the robustness of the edge states \cite{KitagawaExp} and help suppress the limitations on strongly-driven systems. 

In this paper, we take another step to show that quantum walks are even better simulators of topological phenomena that was perceived before. In particular, we address two issues; a) simulation of all types of boundary and edge states and b) high control over the simulated topological phases, boundary and edge states, hence phase transitions. In previous studies, the emphasis was on the simulation of topological phases but the boundary states were neglected \cite{Kitagawa,Kitagawa2012,Asboth,Obuse,Chen}. Here, we use step-dependent coins in one- and two-dimensional quantum walk with two protocols of simple-step (shift-coin operator) and split-step (shift-coin-shift-coin operator) to simulate trivial and nontrivial phases, boundary and edge states. We show that one-dimensional quantum walks simulate all types of boundary and edge states in addition to all types of topological phases of BDI family. Furthermore, the step-dependent coins enable one to readily control the simulations by just choosing a proper number of steps. This controlling factor enables us to determine the presence or absence of boundary states and topological phases, their types and where they should be observed. In addition, we report on emergences of cell-like structure in the simulated topological phases and boundary (edge) states under certain conditions. Each cell contains all types of boundary states and topological phases of BDI family.

The structure of the paper is as follows. First, we outline the relation between unitary operation of the quantum walk and Hamiltonian governing topological phenomena present in the state of a matter. Next, we use simple-step protocol for quantum walk in one-dimensional position space and show that specific types of topological phases and boundary states could be simulated by it. Then, we modify the simple-step protocol to split-step and show that this modification enables us to simulate all types of boundary states and topological phases. We use position-dependent rotation angle to confirm simulation of all types of edge states with our setup. In addition, we generalize our study to two-dimensional position space and investigate the possibility of simulation of different boundary states for two protocols simple- and split-step. The paper is concluded by closing remarks.

\section{General scheme} \label{General}

Formally, the quantum walk with step-independent coin is generated by a successive application of an unitary protocol upon an initial state of a walker, i.e. a photon or an ion,
\begin{eqnarray}
\ketm{\phi}_{fin} & = & \ketm{\phi}_{T}
\;=\; \widehat{U}^{\:T}\: \ketm{\phi}_{int}\; \, \label{protocol}.
\end{eqnarray}

For quantum walk with step-dependent coins, the final state of the walker is achieved through slightly different successive application of an unitary protocol on similar walkers
\begin{eqnarray}
\ketm{\phi}_{fin} & = & \ketm{\phi}_{T}
\;=\; \prod\limits_{m=1}^{T}\widehat{U}(m) \ketm{\phi}_{int}\; \, \label{protocol0},
\end{eqnarray}
where in each step, the coin operator changes.

The (initial) state of the walker is given by walker's internal degrees of freedom, i.e. polarization \cite{Barkhofen2018,Schreiber,Bian} or spin \cite{Karski,Zahringer,Dadras}, and external degrees of freedom, i.e.  optical lattice \cite{Karski} or angular momentum \cite{Dadras,Cardano2015}.

The protocol of the quantum walk consists of coin and shift operators, and it describes in which sequence the coin and shift operators act upon the state of the walker. While the coin operators change the internal degrees of freedom, the shift operators modify the external degrees of freedom based on walker's internal degrees of freedom (internal states). The numbers of the coin and shift operators determine the type of protocol. In literature, two types of protocols have been used for the quantum walk: a) simple-step which has one coin and one shift operator \cite{Panahiyan2019}, b) split-step which includes at least two different coin and two different shift operators \cite{Kitagawa}. In this paper, we use both of these protocols to simulate topological phases, boundary and edge states.

The protocols of the quantum walk are quite similar to Su-Schrieffer-Heeger model which describes electrons (or fermions) that hop on a one-dimensional lattice with staggered hopping amplitudes \cite{Su,Asboth2016}. In addition, since the quantum walk is obtained due to repeated application of a unitary operator, we can describe the evolution of the quantum walk in the framework Floquet theory. Therefore, we can map the evolution of the quantum walk to a stroboscopic evolution under an effective Hamiltonian \cite{Kitagawa,Panahiyan2019}

\begin{eqnarray}
\widehat{H}(k)& = &i \ln\widehat{U}(k)= E(k) \boldsymbol n(k)\cdot \boldsymbol \sigma,  \label{Hamiltonian}
\end{eqnarray}
where $E(k)$ is the energy dispersion, $\boldsymbol \sigma$ are Pauli matrices and $\boldsymbol n(k)$ defines the eigenstates of the energy. With this mapping, we can connect the energy of the Hamiltonian to the eigenvalues of the protocol of quantum walk through
   
\begin{equation} 
\lambda(k)=e^{-i E(k)}, \label{enen}
\end{equation}
where $\lambda(k)$ is the eigenvalue of $\widehat{U}$. The Hamiltonian and its energy are used to investigate topological phases and boundary (edge) states. Therefore, due to one to one correspondence between Hamiltonian (energy) and the protocol of the quantum walk (eigenvalues of $\widehat{U}$), one can use the quantum walk to simulate and investigate the topological phases, boundary (edge) states, and topological phase transitions. The properties of the topological phases, boundary (edge) states can be characterized by the group velocity of the energy eigenstates \cite{Kitagawa,Panahiyan2019}
 
\begin{eqnarray}
V(k)& = & \frac{d E(k)}{dk}. \label{groupv}
\end{eqnarray} 

In this paper, we focus on topological phases with chiral symmetry. Therefore, there is an unitary and Hermitian operator, $\widehat{\Gamma}$, that satisfies two conditions of $\widehat{\Gamma}^2=I$ and $\widehat{\Gamma}\widehat{H}(k)\widehat{\Gamma}=-\widehat{H}(k)$. Presence of chiral symmetry limits the $\boldsymbol n(k)$ to lie on a great circle of the Bloch sphere and winding number is interpreted as a topological invariant. The winding number is defined as number of times $\boldsymbol n(k)$ winds around the origin as $k$ transverses through the first Brillouin zone, $[-\pi,\pi]$. One can obtain the chiral symmetry operator, $\widehat{\Gamma}$, as 

\begin{eqnarray}
\widehat{\Gamma} & = &\boldsymbol A \cdot \boldsymbol\sigma, \label{gamma}
\end{eqnarray}
where $\boldsymbol A$ is a vector labeling a point on the Bloch sphere and perpendicular to $\boldsymbol n(k)$ for all $k$. 

We are interested in the quantum walks that have one and two external degrees of freedom (position space) with two internal degrees of freedom. For the case of one-dimensional position space, the Hilbert space of the walk is given by tensor product of two subspaces, $\mathcal{H} \:\equiv\: \mathcal{H}_{P} \otimes \mathcal{H}_{C}$ where the position Hilbert space ($\mathcal{H}_{P}$) and the coin Hilbert space ($\mathcal{H}_{C}$) are spanned by $\{ \ketm{i}_{P}: i\in \mathbb{Z}\}$ and $\{ \ketm{0},\: \ketm{1} \}$, respectively. In case of two external degrees of freedom, the position Hilbert space is spanned by $\{ \ketm{i,j}_{P}: i,j \in \mathbb{Z}\}$. 

The protocols that we will consider for quantum walks are Hermitian and their determinants are
$1$. Therefore, the Hamiltonians associating to the protocols are traceless which lead to the energy bands having symmetry of $E(k)=E(-k)$ and the energy's value traverses $[-\pi,\pi]$. Since there are two internal states, there will be two bands of energy. The topological phases are where these two energy bands are gapped. The boundary states are located between topological phases and they are determined by gapless energy bands. The topological phase transitions take place at the boundary states and over them. 

In general, there are three ways in which the energy bands can close their gap \cite{Chiu}. This essentially means three types of boundary (and edge) states. The first type is known as Dirac cone where the energy bands are linear functions of the momentum, $k$, and linearly close the gap. The second type is Fermi arc where the gapless energy bands close the gap nonlinearly. Finally, we have flat bands. In the flat bands, energy is constant and independent of momentum. It should be noted that gap-closers happen at $E=0$ and $\pm \pi$. In what follows, we are mainly interested in edge states that manifests at boundary states. This indicates gapless edge states. 

\section{simple-step Quantum walk} \label{simple}

In this section, we employ the simple-step protocol for the quantum walk (simple-step quantum walk) to simulate topological phases, boundary states and topological phase transitions. 

\subsection{General details}

The simple-step protocol consists of one coin and one shift operator 

\begin{eqnarray}
\widehat{U} & = & \widehat{S}_{\uparrow \downarrow} \widehat{C}_{\theta}  \label{protocol1},
\end{eqnarray}
which shows that one step of the quantum walk comprises rotation of internal states with $\widehat{C}_{\theta}$ and displacement of its position with $\widehat{S}_{\uparrow \downarrow}$. The coin operator is step-dependent and considered as \cite{Panahiyan2018,Dhar}

\begin{eqnarray}
\widehat{C}_{\theta}= e^{-\frac{i T \theta}{2}\sigma_{y}}  \label{coin1},
\end{eqnarray}
where $\theta$ is the rotation angle spanning $[0,2\pi]$, $T$ is the number of steps characterizing step-dependency of the coin operator. The shift operator is 

\begin{equation}
\widehat{S}_{\uparrow \downarrow} = \ketm{\uparrow} \bram{\uparrow} \otimes \sum_{x} \ketm{x+1} \bram{x}+
\ketm{\downarrow} \bram{\downarrow} \otimes \sum_{x} \ketm{x-1} \bram{x}\, . \label{shift1}
\end{equation}

We can use Discrete Fourier Transformation, $\ketm{k}=\sum_{x}e^{-\frac{i k x}{2}}\ketm{x}$, to rewrite the shift operator in diagonalized form of

\begin{equation}
\widehat{S}_{\uparrow \downarrow}= e^{ik \sigma_{z}}. \label{shift11}
\end{equation}

It is a matter of calculation to find the eigenvalues of the $\widehat{U}$ in Eq. \eqref{protocol1} as

\begin{equation}
	\lambda(k)= \cos(k) \cos(\frac{T \theta}{2}) \pm \sqrt{[ \cos(k) \cos(\frac{T \theta}{2})]^2-1},
\end{equation}
which by using Eq. \eqref{enen}, we can obtain the energy as \cite{Panahiyan2019} 

\begin{equation}
E(k)= \pm\cos^{-1} \bigg[ \cos(\frac{T \theta}{2}) \cos(k) \bigg]. \label{energy1}
\end{equation}

The positive and negative branches of energy correspond to the two bands of energy. These two bands of energy are gapped and their gap could close only at $E=0$ and $\pm \pi$. The gap-closer happens at $E=\pi$ if $\theta$ admits \cite{Panahiyan2019}

\begin{equation}
	\theta_{E=\pi}=\frac{\pm 2\cos ^{-1} [-\sec (k)]+4\pi c}{T}=\left\{
	\begin{array}{cc} 
		\frac{4\pi c}{T} & k=-\pi
		\\ [0.2cm]
		\frac{\pm 2\pi+4\pi c}{T} & k=0
		\\[0.2cm]
		\frac{4\pi c}{T} & k=\pi    
	\end{array}  
	\right.  , 
\end{equation}
where $c$ is an integer. Similarly, for gap-closer with $E=0$, we have \cite{Panahiyan2019}

\begin{equation}
	\theta_{E=0}=\frac{\pm 2\cos ^{-1} [\sec (k)]+4\pi c}{T}=\left\{
	\begin{array}{cc} 
		\frac{\pm 2\pi+4\pi c}{T} & k=-\pi
		\\ [0.2cm]
		\frac{4\pi c}{T} & k=0
		\\[0.2cm]
		\frac{\pm 2\pi+4\pi c}{T} & k=\pi    
	\end{array}  
	\right.  . 
\end{equation}

On the other hand, we can find cases where energy is independent of momentum and constant, $E=\frac{\pi}{2}$. This takes place for 

\begin{eqnarray}
	\theta_{E=cte=\pi/2}=\frac { 4 \pi  c \pm \pi } {T}.
\end{eqnarray}

It should be noted that $c$ must be chosen in a way that obtained solutions are within $[0,2\pi]$. The flat bands represent existences of degenerate localized states. In the next step, we find $\boldsymbol n(k)$ as 

\begin{equation}
	\boldsymbol n(k) =\frac{\left(\cos (\frac{T\theta}{2}) \sin (k),\sin (\frac{T\theta}{2}) \cos (k),- \cos (\frac{T\theta}{2}) \sin (k) \right)}{\sin E(k)}, \label{n}
\end{equation}
which confirms that when the energy bands close their gap, $\boldsymbol n(k)$ becomes ill-defined. Therefore, the boundary states are characterized by ill-defined $\boldsymbol n(k)$. Since the winding number is the topological invariant, we can conclude that the winding number is also ill-defined at boundary states which is in agreement with our expectation.  

We find the group velocity using Eqs. \eqref{groupv} and \eqref{energy1}

\begin{equation}
V(k)= \pm \frac{ \cos(\frac{T \theta}{2})  \sin(k)}{ \sqrt{1- [\cos(\frac{T \theta}{2}) \cos(k)]^2}}, \label{groupv1}
\end{equation}
which has symmetry of $V(k)=-V(-k)$. Obtained group velocity becomes ill-defined at roots ($\theta'$) of denominator of the group velocity \cite{Panahiyan2019}

\begin{eqnarray}
	\theta'=\left\{
	\begin{array}{cc} 
		\frac{\pm 2\cos ^{-1} [-\sec (k)]+4\pi c}{T}
		\\[0.2cm]
		\frac{\pm 2\cos ^{-1} [\sec (k)]+4\pi c}{T}   
	\end{array}  
	\right.  , 
\end{eqnarray} 
which are coincidence with the calculated points where the energy bands are gapless. 

Finally, in order to obtain the chiral symmetry operator \eqref{gamma}, we find $\boldsymbol A=\left(\cos (\frac{T\theta}{2}),0, \sin (\frac{T\theta}{2})\right)$ which is perpendicular to $\boldsymbol n(k)$ for arbitrary $k$ and gives us the chiral symmetry operator in form of

\begin{equation}
\widehat{\Gamma}= \begin{pmatrix}
\sin(\frac{T \theta}{2}) &   \cos(\frac{T \theta}{2}) \vspace{0,25cm}\\  
\cos(\frac{T \theta}{2}) &  -\sin(\frac{T \theta}{2})
\end{pmatrix},
\end{equation}
which admits both conditions of $\widehat{\Gamma}^2=I$ and $\widehat{\Gamma}\widehat{H}(k)\widehat{\Gamma}=-\widehat{H}(k)$. 

To find the topological invariant (winding number), we use \cite{Cardano2017,Asboth2016} 

\begin{eqnarray}
\gamma=\int_{- \pi}^{\pi} \bigg (\boldsymbol n(k) \times \frac{\partial \boldsymbol n(k)}{\partial k} \bigg) \cdot \boldsymbol A \frac{dk}{2 \pi}. \label{gamma1}
\end{eqnarray} 

It is a matter of calculation to obtain the winding number as  

\begin{eqnarray}
\gamma=-\sin(\frac{T \theta}{2}) \sqrt{\csc(\frac{T \theta}{2})}.
\end{eqnarray}

\begin{figure*}[htb]
	\centering
	{\begin{tabular}[b]{cc}%
			\sidesubfloat[\label{E-Simple}]{\includegraphics[width=1\linewidth]{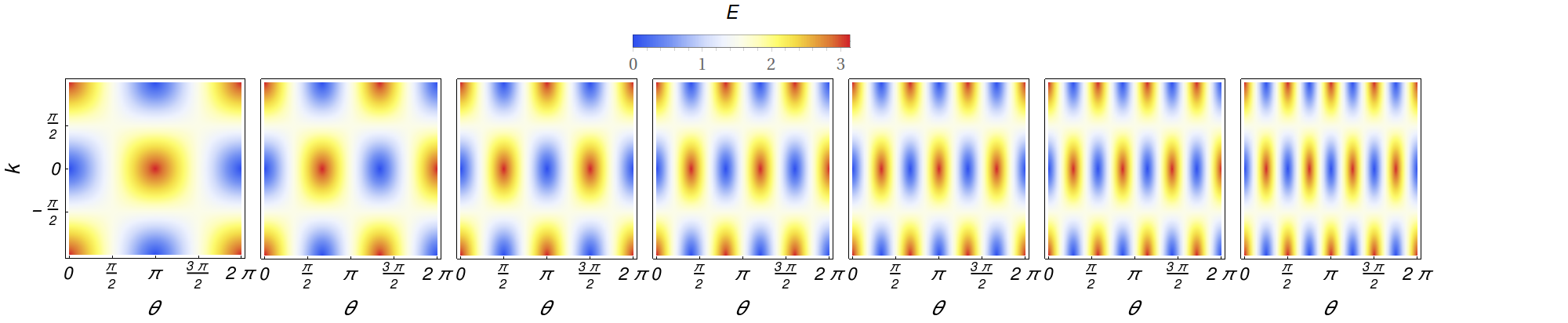}}\\[0.0001cm]
			\sidesubfloat[\label{V-Simple}]{\includegraphics[width=1\linewidth]{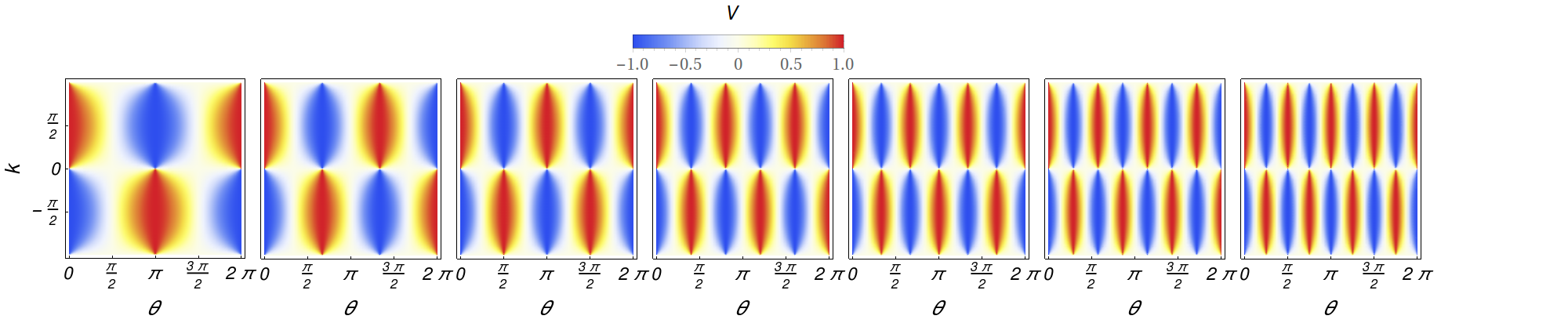}}\\[0.0001cm]
	\end{tabular}}				
	\caption{Simple-step quantum walk (one-dimensional): Modification of energy (a) and group velocity (b) (both positive branches) as functions of momentum and rotation angle $\theta$ for subsequent steps of $T=2,...,8$ from left to right. In (a), we observe only Dirac cone boundary states. As number of steps increases, the distance between two rotation angles both admitting gapless energy bands decreases. It is evident that by tuning the number of steps, we can engineer the place of boundary states and the type of topological phases for each rotation angle. In (b), we see that the group velocity at gapless points is ill-defined. In contrast, the group velocity is constant around gapless points and its signature swaps from positive to negative and vice versa depending on the energy of gapless point.} \label{Fig1}
\end{figure*}	

\subsection{Results}

The major consequence of the step-dependent coin is step-dependent nature for Hamiltonian, energy bands, group velocity, $\boldsymbol n(k)$ and even chiral symmetry operator. This is in complete contrast to simple-step quantum walk with step-independent coin where the step number of the quantum walk has no effect on these properties. 

The step-dependent nature of energy bands indicates that through the quantum walk, the bands of energy associating to each rotation angle will change. Therefore, we can use the step number as a controlling factor and by tuning it, we can decide if the energy bands for specific rotation angle close their gaps, or in which topological phases they should be located (see Fig. \ref{Fig1}a). Consequently, with step-dependent coin, we can engineer the number of topological phases, their sizes, boundary states and their places (phase transition points). This provides us with outstanding level of control over simulations. 

In our study, we established that boundary states are where $\boldsymbol n(k)$ and group velocity become ill-defined. Careful examination shows that $n_{z}=-|V(k)|$ which indicates that group velocity is bounded within $[-1,1]$. Over the phase transition point, the signature of the group velocity suddenly swaps from negative to positive or vice versa (see Fig. \ref{Fig1}b). For positive branch of group velocity, the negative to positive swap happens if energy of the closing gap is $E=0$ and the opposite happens for $E= \pi$. Therefore, the sudden changes in the signature of the group velocity are signs of phase transitions, hence existence of boundary states. 

Here, we observe non-trivial topological phases with topological invariants $\pm 1$ \cite{Panahiyan2019}. The boundary states are located between two topological phases with two different topological invariants. Therefore, we have phase transitions between two different phases. The type of boundary states observed for the simple-step quantum walk is Dirac cone. In fact, the obtained gapless points confirm that it is not possible to observe any other type of boundary state except Dirac cone one (see Fig. \ref{Fig1}a). Therefore, the simple-step quantum walk can only simulate Dirac cone boundary states and two non-trivial topological phases with topological invariants $\pm 1$. To accomplish the simulation of the different types of topological phases and boundary states, we should use the split-step protocol for the quantum walk (split-step quantum walk).

\begin{figure*}[htb]
	\centering
	{\begin{tabular}[b]{cc}%
			\sidesubfloat[\label{E-Split1}]{\includegraphics[width=1\linewidth]{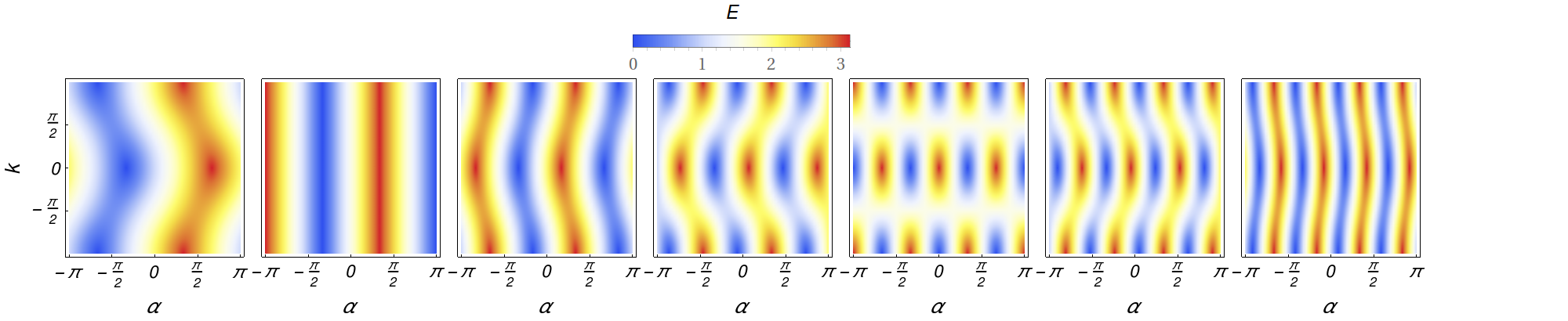}}\\[0.0001cm]
			\sidesubfloat[\label{V-Split1}]{\includegraphics[width=1\linewidth]{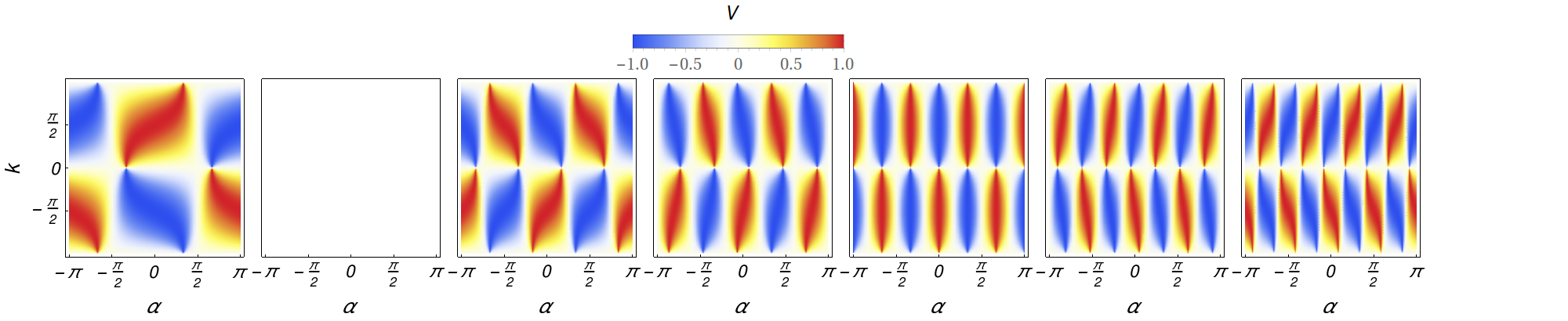}}\\[0.0001cm]
			\sidesubfloat[]{\includegraphics[width=1\textwidth]{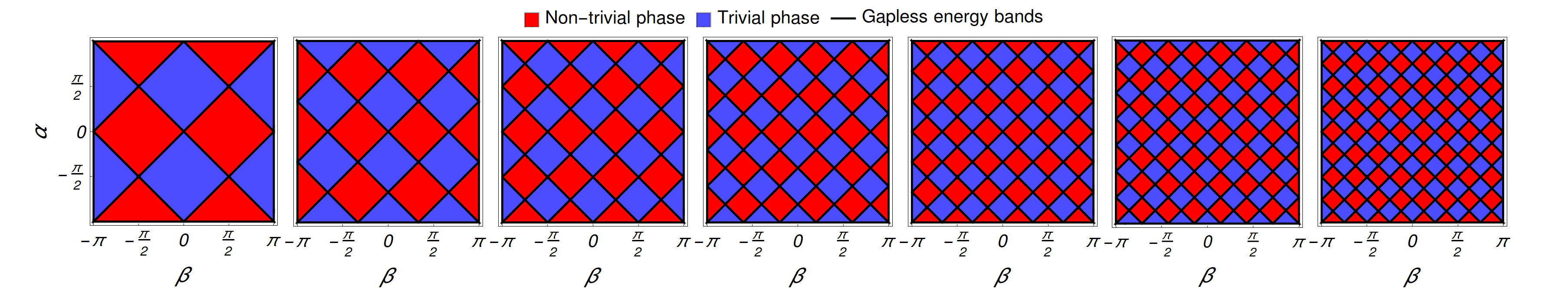}}	
	\end{tabular}}				
	\caption{Split-step quantum walk (one-dimensional): Modification of energy (a) and group velocity (b) (positive branches) as functions of momentum and rotation angle $\alpha$ ($\beta=\pi/3$) and $C$ (c) for subsequent steps of $T=2,...,8$ from left to right. In (a), only one type of the boundary state is present at each step. A network of flat bands occurs in third step where energy bands are independent of $k$. Correspondingly, the group velocity for network of flat bands is zero (b). In (c), the trivial (blue areas with $C>1$) and non-trivial (red areas with $C<1$) phases are marked using $C$. For $C=1$, the energy bands close their gap. The places of topological phases and boundary states, and their numbers are modified step-dependently. This makes the step-number as an engineering mean.} \label{Fig2}
\end{figure*}	
\begin{figure*}[htb]
	\centering
	{\begin{tabular}[b]{cc}%
			\sidesubfloat[\label{E-Split}]{\includegraphics[width=1\linewidth]{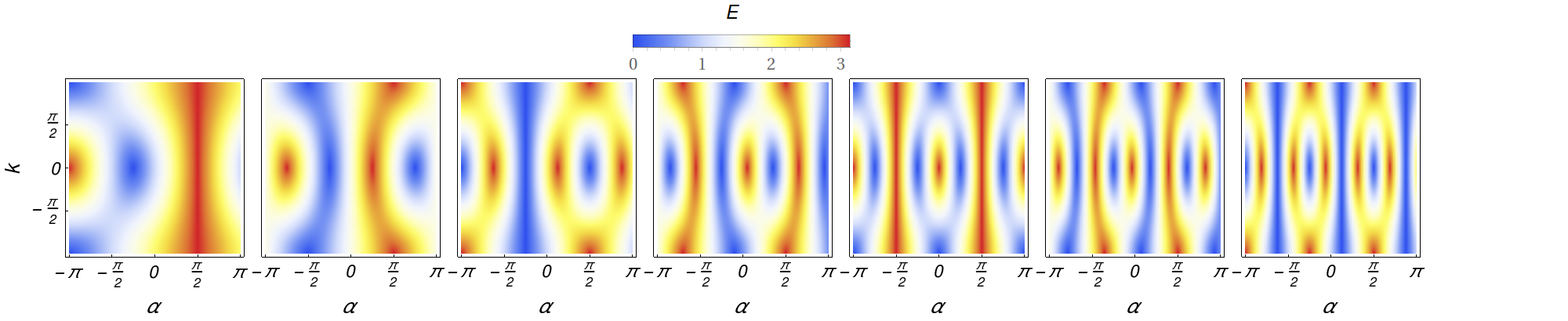}}\\[0.0001cm]
			\sidesubfloat[\label{V-Split}]{\includegraphics[width=1\linewidth]{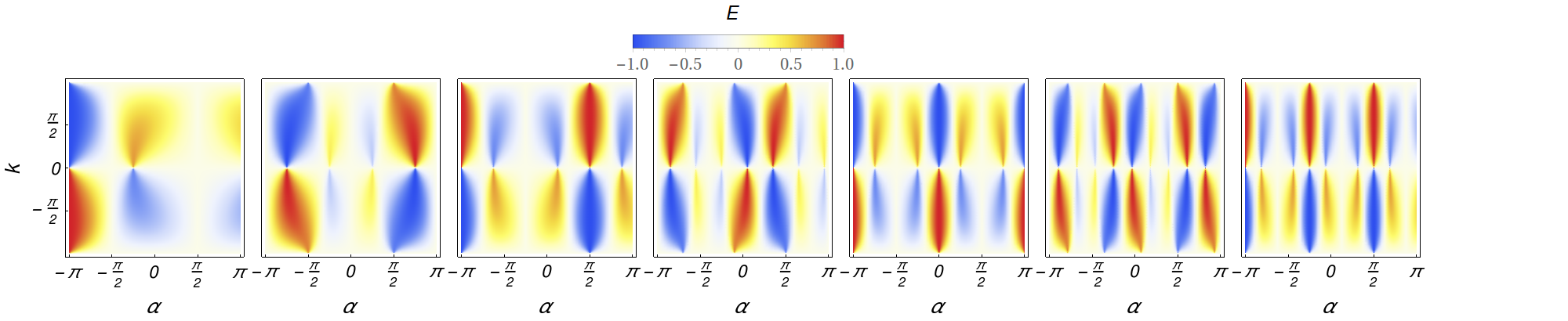}}\\[0.001cm]
			\sidesubfloat[]{\includegraphics[width=1\textwidth]{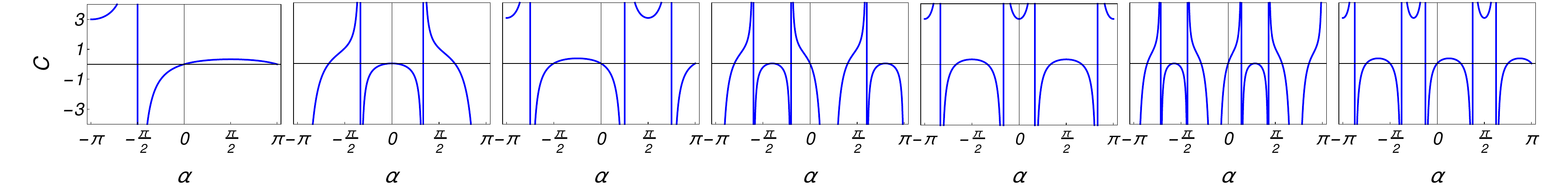}}				
	\end{tabular}}				
	\caption{Split-step quantum walk (one-dimensional): Modification of energy (a) and group velocity (b) (positive branches) as functions of momentum and $C$ (c) with $\alpha$ ($\beta=(\alpha+\pi)/3$) for subsequent steps of $T=2,...,8$ from left to right. In (a), all three types of boundary states are emerged for specific steps and formation of cell-like structures is observed. The cells are marked by two flat bands (acting as cells' walls) and a Dirac cone located between two Fermi arcs. Therefore, all types of boundary states are presented in the same step. In (b), we see that group velocity is ill-defined at Dirac cone and Fermi arc gapless energy bands while it is zero for flat bands. In (c), the trivial ($C>1$), non-trivial ($C<1$) phases and gapless energy bands are marked using $C$. The Dirac cones are recognized by non-zero minimums, flat bands by non-zero maximums and Fermi arcs by divergent points. There is another type of Fermi arc which occurs if $C=1$. The place of the boundary states and topological phases change step-dependently.} \label{Fig3}
\end{figure*}	
\begin{figure}[!htbp]
	\centering
	{\begin{tabular}[b]{ccc}
			\subfloat[]{\includegraphics[width=0.52\textwidth]{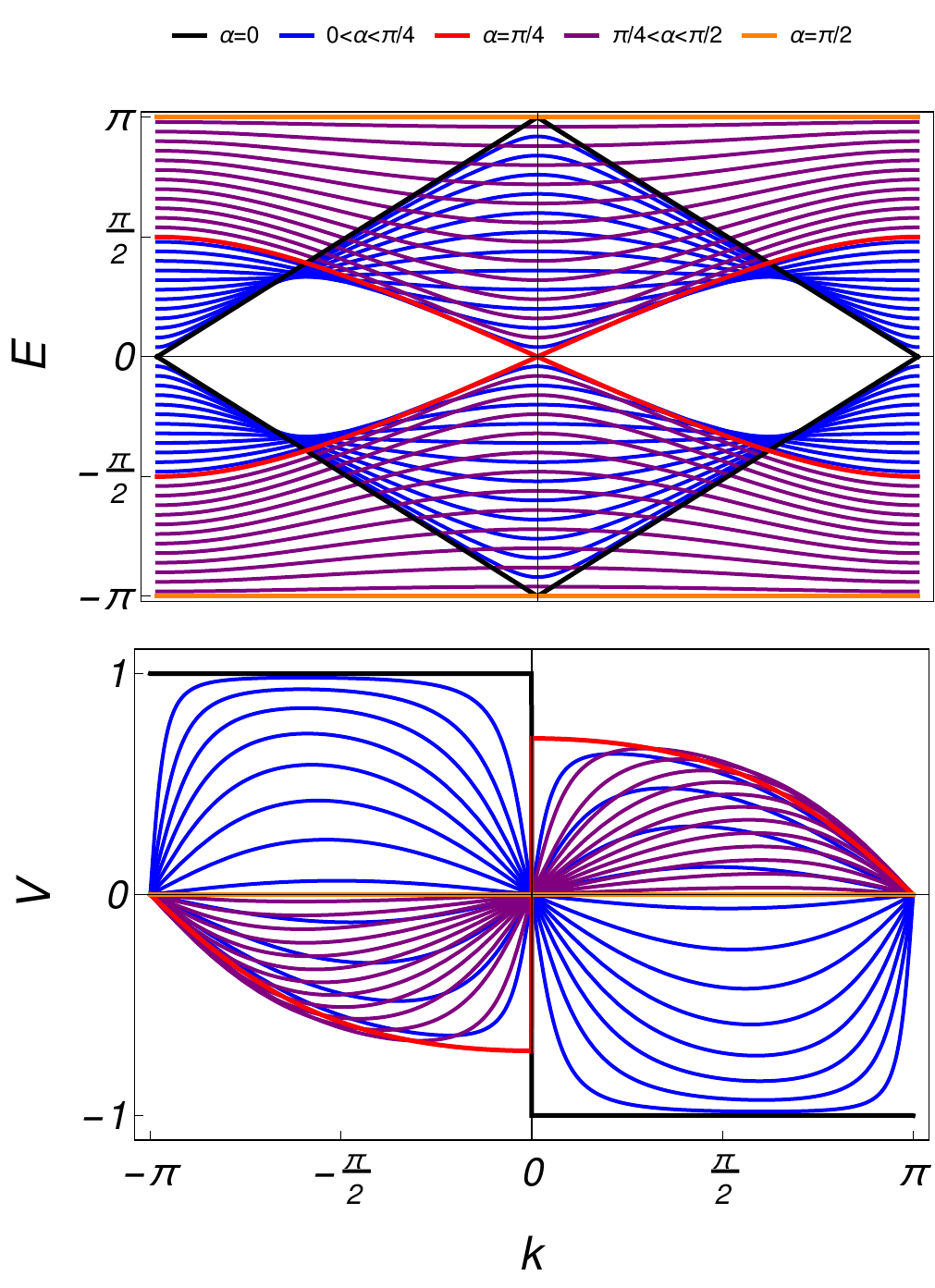}}
			\subfloat[]{\includegraphics[width=0.52\textwidth]{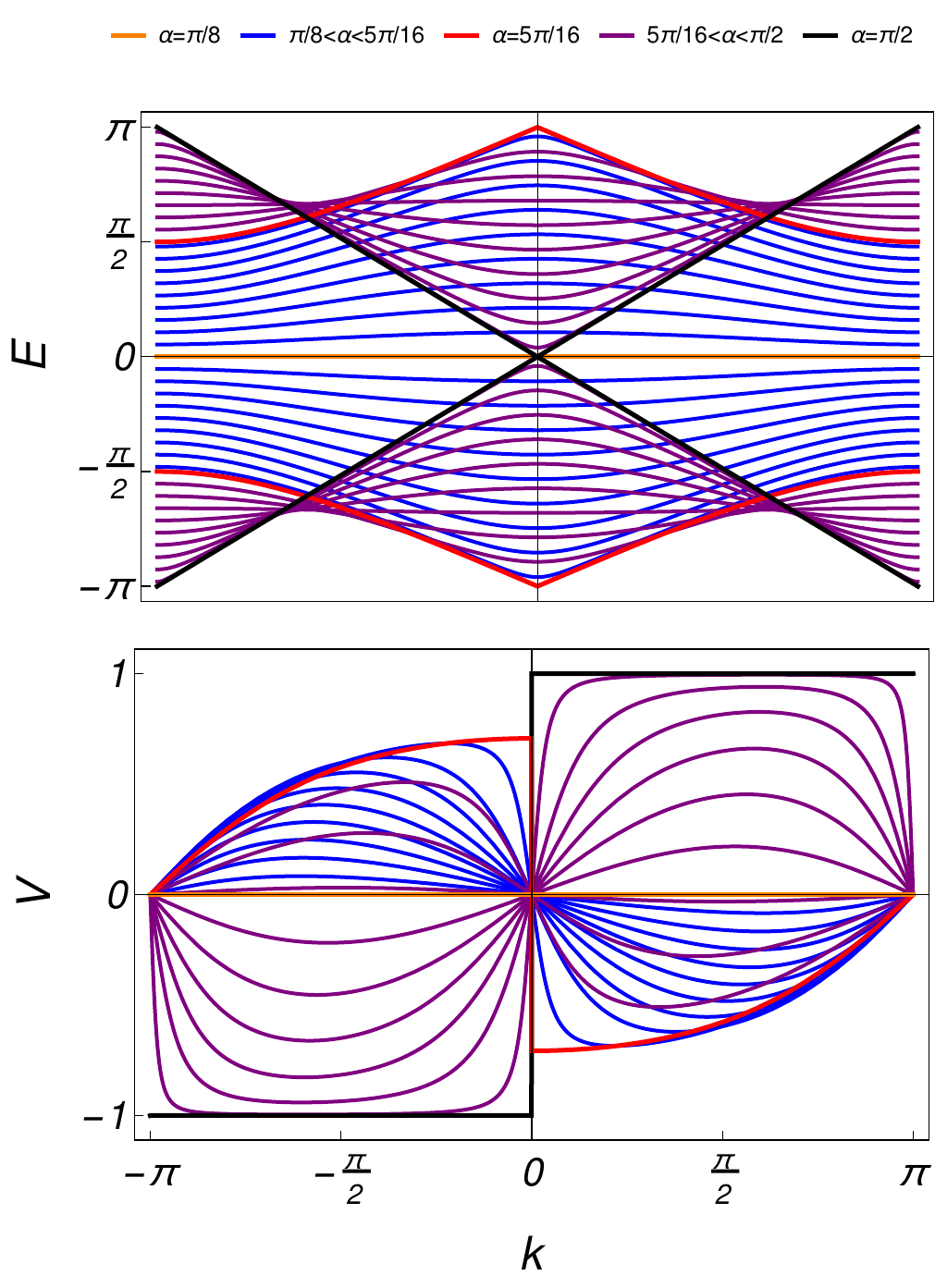}}				
	\end{tabular}} 	
	\caption{Energy (upper panels) and group velocity (lower panels) (positive branches) for $6th$ (a) and $8th$ (b) steps with $\beta=(\alpha+\pi)/3$. The Dirac cones show linear dispersion. The Fermi arcs have nonlinear dispersive behaviors whereas the flat bands are dispersionless. As we span $\alpha$ through $[0,\pi]$, the energy bands close their gap through different boundary states.}	\label{Fig4}
\end{figure}	
\begin{figure*}[!htbp]
	\centering
	{\begin{tabular}[b]{ccc}
			\sidesubfloat[]{\includegraphics[width=1\linewidth]{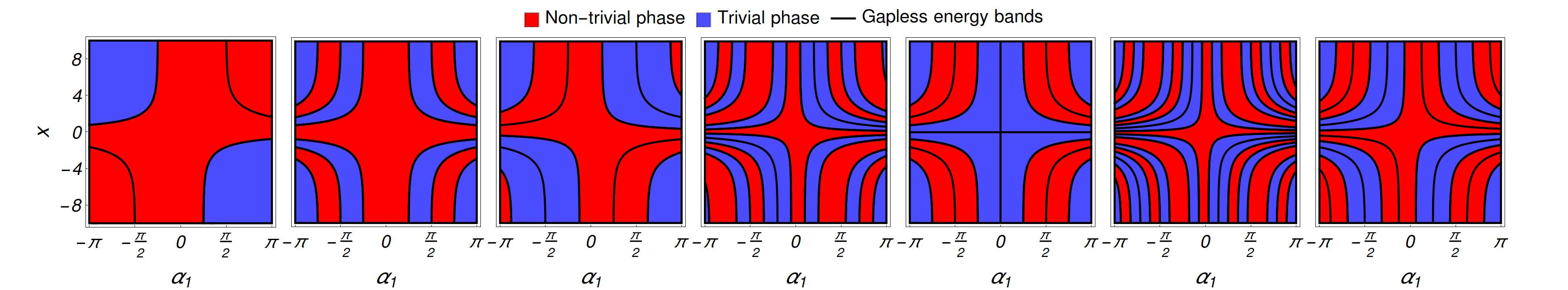}}\\
			\sidesubfloat[]{\includegraphics[width=1\linewidth]{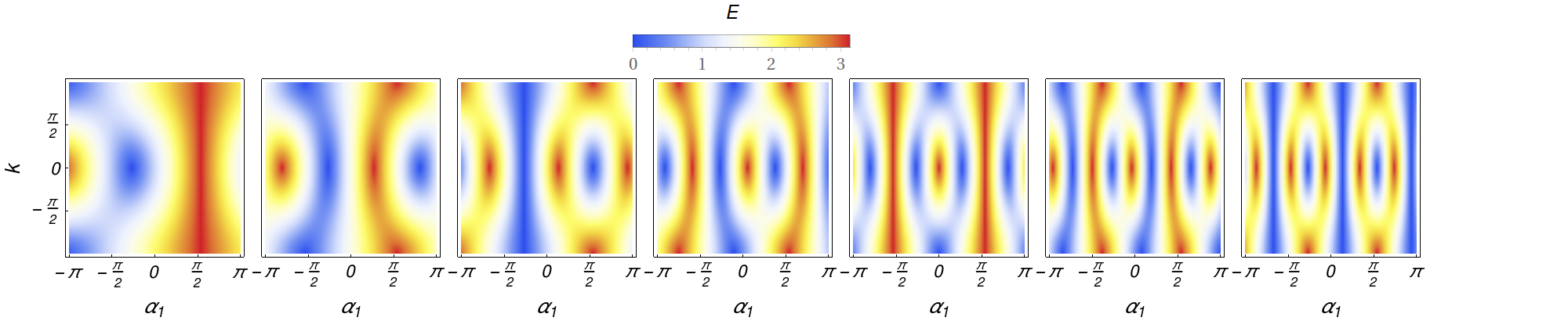}}\\
			\sidesubfloat[]{\includegraphics[width=1\linewidth]{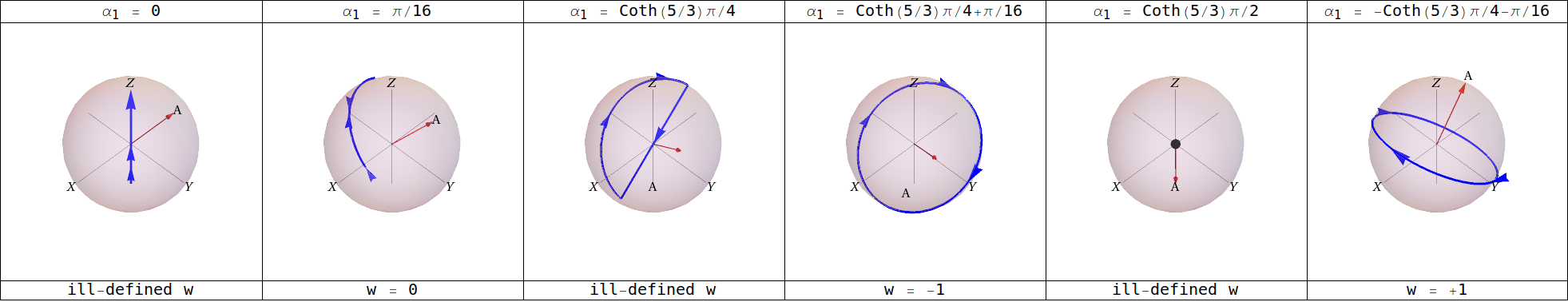}}					
	\end{tabular}} 	
	\caption{Split-step quantum walk (one-dimensional): In (a), the trivial (blue areas) and non-trivial (red areas) phases are marked using $C$ for subsequent steps of $T=2,...,8$ from left to right with $\beta=(\alpha+\pi)/3$ and $\alpha=\alpha_{1}\tanh(x/3)$. In case of $C>1$, the topological phases are trivial while for $C<1$, the topological phases are non-trivial ones. The energy bands close their gap if $C=1$. These gapless energy bands host gapless edge states. As the step number of inhomogeneous quantum walk (due to position dependent coins) increases, the place of edge states and their number vary which indicate that number of topological phases and their corresponding edge states vary as step-number of the quantum walk increases. In (b), modification of energy as a function of momentum and rotation angle $\alpha\alpha_{1}$ with $x=5$ is presented. We observe that cell-like structure reported before emerges for inhomogeneous quantum walk as well. This indicates that each cell contains all types of edge state. In (d), the winding number is depicted through $\boldsymbol n(k)$ (blue curves) as $k$ traverses the first Brillouin zone in a plane orthogonal to $\boldsymbol n(k)$ (red vector) for $T=6$. Evidently, the inhomogeneous quantum walk creates set of boundaries in which the phases on each side are topologically distinct and the BDI family of topological phases is simulated by it.}	\label{Fig4E1}
\end{figure*}	

\section{split-step Quantum walk}  \label{split}

In this section, we modify the simple-step protocol of the quantum walk into a split-step one and investigate the types of topological phases and boundary states. Additionally, we use position-dependent rotation angle to simulate different types of edge state via split-step quantum walk. 

\subsection{General details}

The split-step protocol has certain flexibility since we can have several different coin and shift operators in this protocol. In this paper, we consider the following split-step protocol  

\begin{eqnarray}
\widehat{U} & = & \widehat{S}_{\uparrow} \widehat{C}_{\alpha} \widehat{S}_{\downarrow} \widehat{C}_{\beta}  \label{protocol2},
\end{eqnarray}
in which, one step of the quantum walk includes rotation of internal states of walker with $\widehat{C}_{\beta}$, displacement of its position with $\widehat{S}_{\downarrow}$, a second rotation of internal states with $\widehat{C}_{\alpha}$ and finally, its displacement with $\widehat{S}_{\uparrow}$. The coin operators are given by 

\begin{eqnarray}
\widehat{C}_{\alpha} & = & e^{-\frac{i T \alpha}{2}\sigma_{y}},  \label{coin2}
\\
\widehat{C}_{\beta} & = & e^{-\frac{i T \beta}{2}\sigma_{y}},  \label{coin3}
\end{eqnarray}
where $\alpha$ and $\beta$ are rotation angles spanning $[-\pi,\pi]$ and $T$ characterizes step-dependency of the coins. The $\widehat{S}_{\uparrow}$ and $\widehat{S}_{\downarrow}$ are shift operators 

\begin{equation}
\widehat{S}_{\uparrow}=\sum_{x}[\ketm{\uparrow} \bram{\uparrow} \otimes \ketm{x+1} \bram{x} + \ketm{\downarrow} \bram{\downarrow} \otimes \ketm{x} \bram{x}\,]  \label{shift2},
\end{equation}

\begin{equation}
\widehat{S}_{\downarrow}=\sum_{x}[\ketm{\uparrow} \bram{\uparrow} \otimes \ketm{x} \bram{x} + \ketm{\downarrow} \bram{\downarrow} \otimes \ketm{x-1} \bram{x}\,]  \label{shift3},
\end{equation}
which by using Discrete Fourier Transformation, we can rewrite them as $\widehat{S}_{\uparrow}=e^{\frac{i k}{2}(\sigma_{z}-1)}$ and $\widehat{S}_{\downarrow}=e^{\frac{i k}{2}(\sigma_{z}+1)}$. We obtain the eigenvalues of the $\widehat{U}$ in Eq. \eqref{protocol2} as 

\begin{eqnarray}
	\lambda(k)= \gamma \pm \sqrt{\gamma^2-1},
\end{eqnarray}
where $\gamma=\cos(k) \cos(\frac{T \alpha}{2})\cos(\frac{T \beta}{2})-\sin(\frac{T \alpha}{2})\sin(\frac{T \beta}{2})$. It is a matter of calculation to find the energy bands as  

\begin{eqnarray}
E(k) & = & \pm\cos^{-1}(\gamma). \label{energy2}
\end{eqnarray}

The energy spans $[-\pi,\pi]$ and the gap between the energy bands can only close at $E=0$ and $E=\pm \pi$. Based on obtained energy bands \eqref{energy2}, there are three possible types for gapless points which are determined by the following conditions:

I) If $\sin(\frac{T \alpha}{2})\sin(\frac{T \beta}{2})=0$, the energy bands close their gap linearly. Therefore, the boundary states are Dirac cone. 

II) If $\cos(k) \cos(\frac{T \alpha}{2})\cos(\frac{T \beta}{2})=0$, energy bands become independent of momentum $k$ and the boundary states are flat bands. 

III) If the two previous conditions are not met, the energy bands close their gaps nonlinearly and the boundary states are Fermi arcs. 

Therefore, depending on these conditions, we have the possibility of simulation of three types of the boundary states. If the rotation angles, $\alpha$ and $\beta$ are independent of each other, only one type of boundary state is observed in each step (see Fig. \ref{Fig2}a). In contrast, if these rotation angles are linearly related to each other, $\beta= s_{1} \alpha + s_{2}$ where $s_{i}$ with $s_{i}$ being real numbers, it is possible to simulate all types of boundary states in a single step (see Fig. \ref{Fig3}a).  

Next, we find $\boldsymbol n(k)$ in the following form 

\begin{equation}
n_{x}(k)  =  \frac{\cos (\frac{T\alpha}{2}) \sin (\frac{T\beta}{2}) \sin (k)}{\sin E(k)}, \notag
\end{equation}
\begin{equation}
n_{y}(k)  =  \frac{\sin (\frac{T\alpha}{2}) \cos (\frac{T\beta}{2}) + \cos (k) \cos (\frac{T\alpha}{2}) \sin (\frac{T\beta}{2}) }{\sin E(k)}, \notag
\end{equation}
\begin{equation}
n_{z}(k)  = - \frac{\cos (\frac{T\alpha}{2}) \cos (\frac{T\beta}{2}) \sin (k)}{\sin E(k)} 
\end{equation} 
which confirms that the gapless points of the energy bands are where $\boldsymbol n(k)$ becomes ill-defined. Therefore, similar to the simple-step case, the boundary states are characterized by ill-defined $\boldsymbol n(k)$ and consequently ill-defined topological invariant. In addition, we obtain the group velocity as 

\begin{eqnarray}
V(k)& = & \pm \frac{ \cos(\frac{T \alpha}{2})\cos(\frac{T \beta}{2})  \sin(k)}{ \sqrt{1- \gamma^2}}. \label{groupv2}
\end{eqnarray} 

Here, the characteristic behavior of the group velocity at the boundary states depends on the type of the boundary state. For gapless energy bands in form of Dirac cone and Fermi arc, the group velocity is ill-defined at the boundary states (see Figs. \ref{Fig2}b and \ref{Fig3}b). Whereas for flat bands boundary states, the group velocity is zero. This is due to fact that energy is independent of momentum, $k$, for flat bands of energy. 

Since we are focused on chiral symmetry for our quantum walks, we find $\boldsymbol A=\left(\cos (\frac{T\beta}{2}),0, \sin (\frac{T\beta}{2})\right)$ and by using it with Eq. \eqref{gamma}, we have the chiral symmetry operator as 

\begin{eqnarray}
\widehat{\Gamma} & = & \begin{pmatrix}
\sin(\frac{T \beta}{2}) &   \cos(\frac{T \beta}{2}) \vspace{0,25cm}\\  
\cos(\frac{T \beta}{2}) &  -\sin(\frac{T \beta}{2})
\end{pmatrix}. \label{chiralsplit}
\end{eqnarray}

In order to investigate the topological structure of the simulated topological phenomena by this quantum walk, one can use the Zak phase. It is a matter of calculation (see section 4.1 in \cite{Puentes} for the method) to find the Zak phase as 

\begin{eqnarray}
C=\frac{\tan (\frac{T \alpha}{2})}{\tan (\frac{T \beta}{2})}.
\end{eqnarray}

If the rotation angles are linearly related to each other, the following holds for obtained Zak phase: The Dirac cones are recognized by non-zero minimums, flat bands by non-zero maximums and Fermi arcs by divergent points. There is another type of Fermi arc which occurs if $C=1$. In case of $C>1$, the topological phases are trivial whereas in case of $C<1$, non-trivial topological phases are observable provided $C$ is not Maximum or minimum that were pointed out. These cases are plotted in Fig. \ref{Fig3}c.

Alternatively, if we have two parameters to characterize our topological structure, one can also use the following expression for Zak phase \cite{Kitagawa}

\begin{eqnarray}
C=|\frac{\tan (\frac{T \alpha}{2})}{\tan (\frac{T \beta}{2})}|,
\end{eqnarray}
in which for $C>1$, the topological phases are trivial while for $C<1$, the topological phases will be non-trivial ones. For $C=1$, the energy bands close their gap. This expression for Zak phase is used in Figs. \ref{Fig2}c and \ref{Fig4E1}a.

\subsection{Results}

Similar to simple-step quantum walk, Hamiltonian, energy bands, group velocity, $\boldsymbol n(k)$ and chiral symmetry operator have step-dependent nature. The chiral symmetry operator and $\boldsymbol A$ are functions of rotation angle $\beta$ and independent of $\alpha$. Chronologically, the coin operator with rotation angle $\beta$ is applied first in the split-step protocol (see \eqref{coin3}). Therefore, the symmetrical properties governing the topological phases are dominated by the first coin operator that is applied on internal states of the walker. 

One of the major differences between simple-step and split-step quantum walks is the presence of $\sin(\frac{T \alpha}{2})\sin(\frac{T \beta}{2})$ term in the energy bands of the split-step quantum walk (compare \eqref{energy1} and \eqref{energy2}). This indeed enriches the topological phases and boundary states that the quantum walk can simulate. In fact, this term is one of the main reasons that the quantum walk with split-step protocol can simulate all types of topological phases, boundary states and later edge states. We further enriched the capability of the quantum walk to simulate topological phases, boundary and edge states by considering step-dependent coin.

As we pointed it out before, the type of boundary state depends on certain conditions governed by the rotation angles, $\alpha$ and $\beta$. In fact, based on these conditions, we can program our simulations to have specific type of gapless energy bands through the whole simulation (see Figs. \ref{Fig2}a and \ref{Fig3}a). On the other hand, the step-dependency of the energy bands enables us to determine number of boundary states or topological phases, and where these boundary states or topological phases should be as rotation angles traverses $[-\pi,\pi]$ (see Figs. \ref{Fig2}c and \ref{Fig3}c). The controlling factor is the number of step. These two properties give us high level of controllability over simulation of topological phenomena with quantum walk. 

The most interesting simulations of topological phases are those when the two rotation angles are linearly related to each other ($\beta= s_{1} \alpha + s_{2}$). First of all, it is possible to simulate all three types of Dirac cone, Fermi arc and flat bands for gapless energy bands in a single step (see Fig. \ref{Fig3}a). This is in contrast to previous case where each boundary state could be simulated in different steps (compare Figs. \ref{Fig2}a and \ref{Fig3}a). The second issue is the coexistence of these three gapless energy bands together in a single step. Finally, the emergences of cell-like topological structures in specific steps (see Fig. \ref{Fig3}a and c for $T=6$ and $\alpha \in [-\pi/2,\pi/2]$). These cell-like structures are characterized by two flat bands which are the cell's walls. The interior part of the cell contains two Fermi arcs, a Dirac cone between them and four phases of trivial and non-trivial ones. The Fermi arc boundary states separate trivial and non-trivial phases whereas the Dirac cone separates two trivial phases from one another. Therefore, a single cell contains all three types of boundary states (topological phase transitions) as well as trivial and non-trivial phases that are observable in BDI family. The emergences of the two flat bands boundary states with the structure that we described point towards the cell-like structure.  

The split-step protocol simulates all types of topological phase observable for BDI family. To prove this, we use the following argument; At the first step, $T=1$, the obtained Hamiltonian, energy bands and other properties yield split-step quantum walk with step-independent coin which was investigated by Kitagawa \textit{et al.} in Ref. \cite{Kitagawa}. Kitagawa \textit{et al.} showed that the split-step quantum walk considered by them simulates all types of topological phase observable for BDI family. Therefore, split-step quantum walk with step-dependent coin in this paper also simulates similar topological phases. 

The gapless energy bands are also characterized by ill-defined $\boldsymbol n(k)$. This is independent of the type of gapless energy bands. Therefore, it is a general property to recognize the boundary states but not distinguish them from one another. On the other hand, the group velocity is related to $\boldsymbol n(k)$ through $n_{z}=-|V(k)|$ which shows that the group velocity spans $[-1,1]$. The behavior of group velocity around gapless points depends on the energy of gapless points and the type of gapless energy bands. 

For Dirac cone gapless energy bands (see Fig. \ref{Fig4}), the group velocity around the boundary state is constant and at gapless point it swaps from positive to negative (for $E=\pi$) or vice versa (for $E=0$). In case of Fermi arc, around the gapless point, the group velocity depends on momentum, $k$, and its signature swaps from positive to negative (for $E=\pi$) or vice versa (for $E=0$) at gapless point. Finally, for flat bands gapless energy bands, group velocity is zero and independent of momentum. These three distinctive behaviors enable us to recognize the type of boundary states.

The group velocity characterizes the motion of a wave packet associating to the walker. Accordingly, the boundary states that are in form of Dirac cones are linearly dispersing \cite{Matsuura2013}. The Fermi arc boundary states show nonlinear dispersive behaviors. In contrast, the flat bands are dispersionless. In fact, the group velocities that are independent of $k$ are known as dispersionless transport and dependency on $k$ indicates that we have dispersive transportation \cite{Groh2016}. Finally, we should point it out that the presence of flat bands signals to existence of a macroscopic number of degenerate localized states \cite{Goldman2011}.

In order to create edge states, we consider position-dependently as a feature of the rotation angle $\alpha$. Such a consideration omits the translationally invariant of the quantum walk but retains the symmetries that were available for the protocol. In Ref. \cite{Kitagawa}, Kitagawa et al. proposed to incorporate the position-dependently in rotation angle through

\begin{eqnarray}
\alpha=\frac{1}{2}(\alpha_{1}+\alpha_{2})+\frac{1}{2}(\alpha_{1}-\alpha_{2})\tanh(x/3).
\end{eqnarray} 
 
Here, we consider $\alpha_{2}=-\alpha_{1}$ which gives us the position-dependent rotation angle as $\alpha=\alpha_{1}\tanh(x/3)$. With $\beta=(\alpha+\pi)/3$, we have an inhomogeneous quantum walk which can simulate topological phases with different topological invariants (see Fig. \ref{Fig4E1}a). We focus only on the edge states with energy $0$ and $\pm \pi$.

Due to step-dependent coins, the capability of engineering topological phenomena through step-number of the quantum walk is present here as well (see Fig. \ref{Fig4E1}a). The resulted phase diagrams and boundary states show that the phases around boundary states can be topologically distinct (see Figs. \ref{Fig4E1}a and c). This indicates that a single localized state with energy $0$ and $\pm \pi$ is present in the boundary state \cite{Kitagawa}. Therefore, we have edge states with energy $0$ and $\pm \pi$. It should be noted that for inhomogenous quantum walk, the position-dependent rotation angle indicates that in real space and at each position, the topological invariant can be locally defined and from one position to another one, the topological invariant could vary (see Figs. \ref{Fig4E1}a and c).

The next issue is whether inhomogeneous quantum walk can simulate cell-like structures that were reported before. To this end, we consider specific position ($x=5$) and investigate the emergence of the cell-like structures. The results in Fig.\ref{Fig4E1}b confirm emergences of these structures for inhomogeneous quantum walk. It should be noted that in these cells, we have simulation of different topological phases of the BDI family. Finally, we notice that edge states observed for simulation could be in forms of Fermi arc, Dirac cone and flat bands. Therefore, we are able to simulate all types of the edge states in one dimension with the quantum walk proposed in this paper. 

To invoke the bulk-boundary correspondence picture for the edge states at the interfaces between two phases (boundary states), one can use the following steps presented in Ref. \cite{Asboth}. We consider set of numbers,  $Q_{0}$ and $Q_{\pi}$, which can be used to describe topological structure of our simulation. First, pick a point in parameter space of phase $A$. Then connect to a point in phase $B$. Finally, start to count the number of times that the gap around $E=0$ and $E=\pm \pi$ closes to obtain $Q_{0}$ and $Q_{\pi}$. This enables us to recognize the differences in both invariants of $Q_{0}$ and $Q_{\pi}$ for gapped phases in different regions of $\alpha_{1}$. 

\section{Generalization to two-dimensional position space}  \label{two dimension}

In this section, we generalize quantum walks to two-dimensional position space, keep its internal states as two and investigate topological phenomena that can be simulated by such generalization. First, we study a two-dimensional quantum walk with simple-step protocol and then split-step protocol. 

\subsection{simple-step Quantum walk}

The protocol of the simple-step quantum walk with two-dimensional position space is given by 
\begin{eqnarray}
\widehat{U} & = & \widehat{S}_{\uparrow \downarrow} \widehat{C}_{\theta}  \label{protocol3},
\end{eqnarray}
where the shift operator includes two other shift operators \cite{Kitagawa2012} 

\begin{equation}
\widehat{S}_{\uparrow \downarrow} = \widehat{S}_{\uparrow \downarrow}(y) \widehat{S}_{\uparrow \downarrow}(x), \label{2shift2}
\end{equation}
in which 
\begin{eqnarray}
\widehat{S}_{\uparrow \downarrow} (x) = &&\ketm{\uparrow} \bram{\uparrow} \otimes \sum_{x,y} \ketm{x+1,y} \bram{x,y}    \notag
\\
&& +\ketm{\downarrow} \bram{\downarrow} \otimes \sum_{x,y} \ketm{x-1,y} \bram{x,y}\, , \label{shift21}
\end{eqnarray}
\begin{eqnarray}
\widehat{S}_{\uparrow \downarrow} (y) = &&\ketm{\uparrow} \bram{\uparrow} \otimes \sum_{x,y} \ketm{x,y+1} \bram{x,y}    \notag
\\
&& +\ketm{\downarrow} \bram{\downarrow} \otimes \sum_{x,y} \ketm{x,y-1} \bram{x,y}\, . \label{shift22}
\end{eqnarray}

A single of step of the walk includes rotation of internal states of the walker with $\widehat{C}_{\theta}$, displacement of its position first in $x$ position space followed by another displacement in $y$ position space. Using Discrete Fourier Transformation, we can find these shift operators as $\widehat{S}_{\uparrow \downarrow} (x)= e^{ik_{x} \sigma_{z}}$ and $\widehat{S}_{\uparrow \downarrow} (y)= e^{ik_{y} \sigma_{z}}$. It is a matter of calculation to find eigenvalues of Eq. \eqref{protocol3} as 

\begin{eqnarray}
	\lambda(k_{x},k_{y}) &=& \cos(k_{x}+k_{y}) \cos(\frac{T \theta}{2}) \pm \notag
	\\
	 & &\sqrt{[ \cos(k_{x}+k_{y}) \cos(\frac{T \theta}{2})]^2-1},
\end{eqnarray}
and consequently, we obtain the energy as 

\begin{eqnarray}
E(k_{x},k_{y})&=& \pm\cos^{-1} \bigg[ \cos(\frac{T \theta}{2}) \cos(k_{x}+k_{y}) \bigg]. \label{energy3}
\end{eqnarray}

\begin{figure*}[htb]
	\centering
	{\begin{tabular}[b]{cc}%
			\sidesubfloat[]{\includegraphics[width=0.3\linewidth]{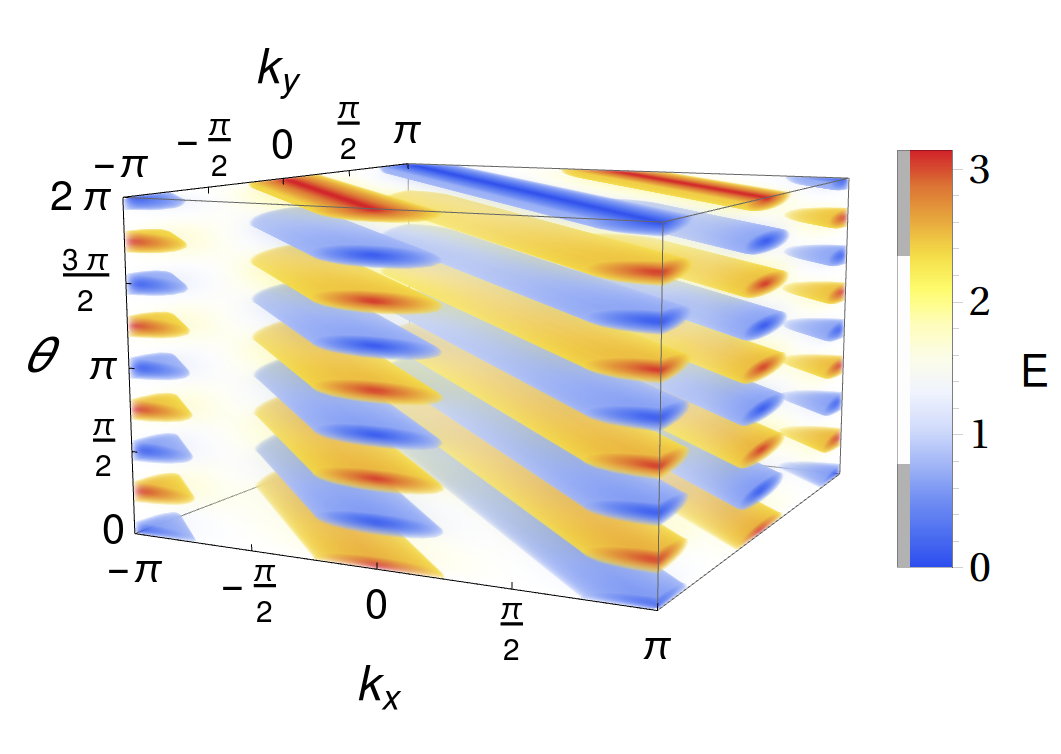}}			
			\sidesubfloat[]{\includegraphics[width=0.65\linewidth]{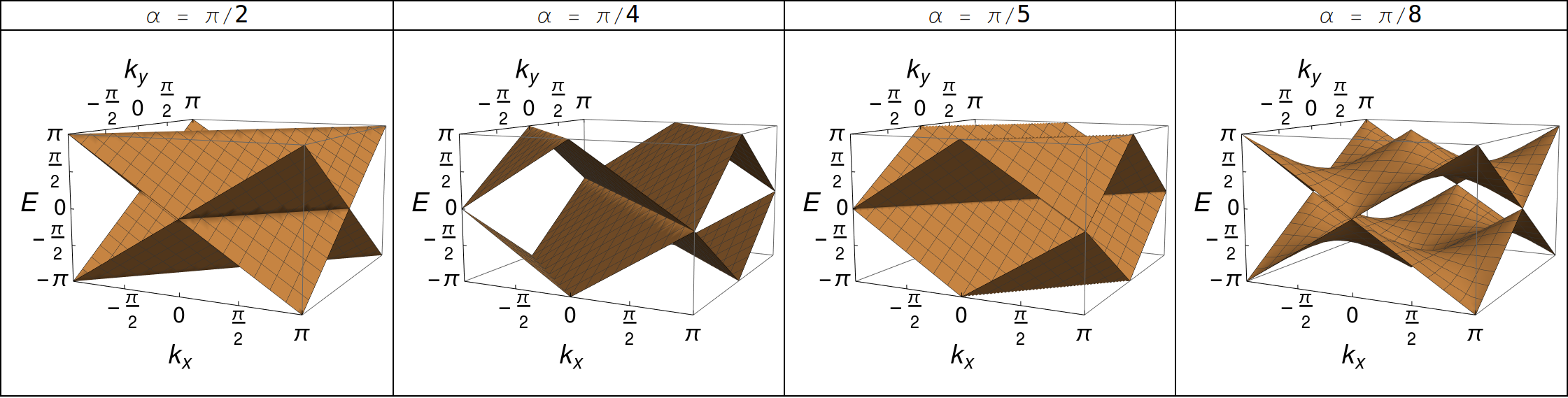}}		
	\end{tabular}}				
	\caption{Simple-step quantum walk (two-dimensional): Energy as a function of rotation angle and momenta for $T=8$. In (a), we observe the existence of several gapless energy bands with similar types. In (b), if we set rotation angle $\theta=\pi/2$ and $\theta=\pi/4$ and consider $k_{y}=-\pi$ and varies $k_{x}$ through the first Brillouin zone, energy bands close their gap linearly (Dirac cones). For variations of both momenta through the first Brillouin zone, gapless energy bands with momenta-dependent flat bands are formed. In case of $\theta=\pi/5$ energy bands are gapped, hence topological phases. For $\theta=\pi/8$, we observe flat bands.} \label{Fig5}
\end{figure*}

We have two bands of energy corresponding to negative and positive branches of energy. $k_{x}$ and $k_{y}$ traverse the first Brillouin zone. Energy's value is limited within $[-\pi,\pi]$ and the gap between the energy bands closes at $E=0$ and $\pm \pi$ (see Figs. \ref{Fig5}a to c) for 

\begin{eqnarray}
	\theta_{E=\pi}&=&\frac{\pm 2\cos ^{-1} [-\sec (k_{x}+k_{y})]+4\pi c}{T}, 
\end{eqnarray}

\begin{eqnarray}
	\theta_{E=0}&=&\frac{\pm 2\cos ^{-1} [\sec (k_{x}+k_{y})]+4\pi c}{T}. 
\end{eqnarray}
where $c$ is an integer. In addition, the energy could be independent of momentum and constant (see Fig. \ref{Fig5}e), $E=\frac{\pi}{2}$ which happens for 

\begin{eqnarray}
	\theta_{E=cte=\pi/2}&=&\frac { 4 \pi  c \pm \pi } {T}.
\end{eqnarray}

Next, we find $\boldsymbol n(k_{x},k_{y})$ as 

\begin{equation}
n_{x}(k_{x},k_{y})  =  \frac{\cos (\frac{T\theta}{2}) \sin (k_{x}+k_{y})}{\sin E(k_{x},k_{y})}, \notag
\end{equation}
\begin{equation}
n_{y}(k_{x},k_{y})  =  \frac{\sin (\frac{T\theta}{2}) \cos (k_{x}+k_{y})}{\sin E(k_{x},k_{y})}, \notag
\end{equation}
\begin{equation}
n_{z}(k_{x},k_{y})  = - \frac{- \cos (\frac{T\theta}{2}) \sin (k_{x}+k_{y})}{\sin E(k_{x},k_{y})}. 
\end{equation} 

It is evident that $\boldsymbol n(k_{x},k_{y})$ becomes ill-defined when the energy bands close their gap. As for the group velocity, since the walker's wave function propagates in two dimensions, we can find two group velocities associating to each position space by 

\begin{equation}
V_{k_{x}}= V_{k_{y}}= \pm \frac{ \cos(\frac{T \theta}{2})  \sin(k_{x}+k_{y})}{ \sqrt{1- [\cos(\frac{T \theta}{2}) \cos(k_{x}+k_{y})]^2}}, \label{groupv3}
\end{equation}
in which we have used $V(k_{x})= \partial E(k_{x},k_{y})/ \partial k_{x}$ and likewise for $V(k_{y})$. Evidently, the group velocity becomes ill-defined at 

\begin{eqnarray}
	\theta'=\left\{
	\begin{array}{cc} 
		\frac{\pm 2\cos ^{-1} [-\sec (k_{x}+k_{y})]+4\pi c}{T}
		\\[0.2cm]
		\frac{\pm 2\cos ^{-1} [\sec (k_{x}+k_{y})]+4\pi c}{T} 
	\end{array}  
	\right.  , 
\end{eqnarray} 
which are the points where the energy bands are gapless. 

It should be noted that $\boldsymbol n(k_{x},k_{y})$ maps a small area on a two dimensional torus to a small area on a Bloch sphere \cite{Kitagawa2012}. The winding number in one-dimensional quantum walk is no longer the topological invariant for two-dimensional quantum walk. In contrast, an integer number known as the Chern number is the topological invariant. The Chern number measures number of times that $\boldsymbol n(k_{x},k_{y})$ covers the Bloch sphere as $k_{x}$ and $k_{y}$ traverse the first Brillouin zone. The Chern number is calculated by \cite{Kitagawa2012}

\begin{eqnarray}
C=\int_{- \pi}^{\pi} \bigg (\frac{\partial \boldsymbol n(k_{x},k_{y})}{\partial k_{x}} \times \frac{\partial \boldsymbol n(k_{x},k_{y})}{\partial k_{y}} \bigg) \cdot \boldsymbol n \frac{d^2k}{4 \pi}, \label{Chern}
\end{eqnarray} 
which is equal to $0$ using obtained $ \boldsymbol n(k_{x},k_{y})$.

\begin{figure*}[htb]
	\centering
	{\begin{tabular}[b]{cc}%
			\sidesubfloat[\label{E-Split-1}]{\includegraphics[width=0.3\linewidth]{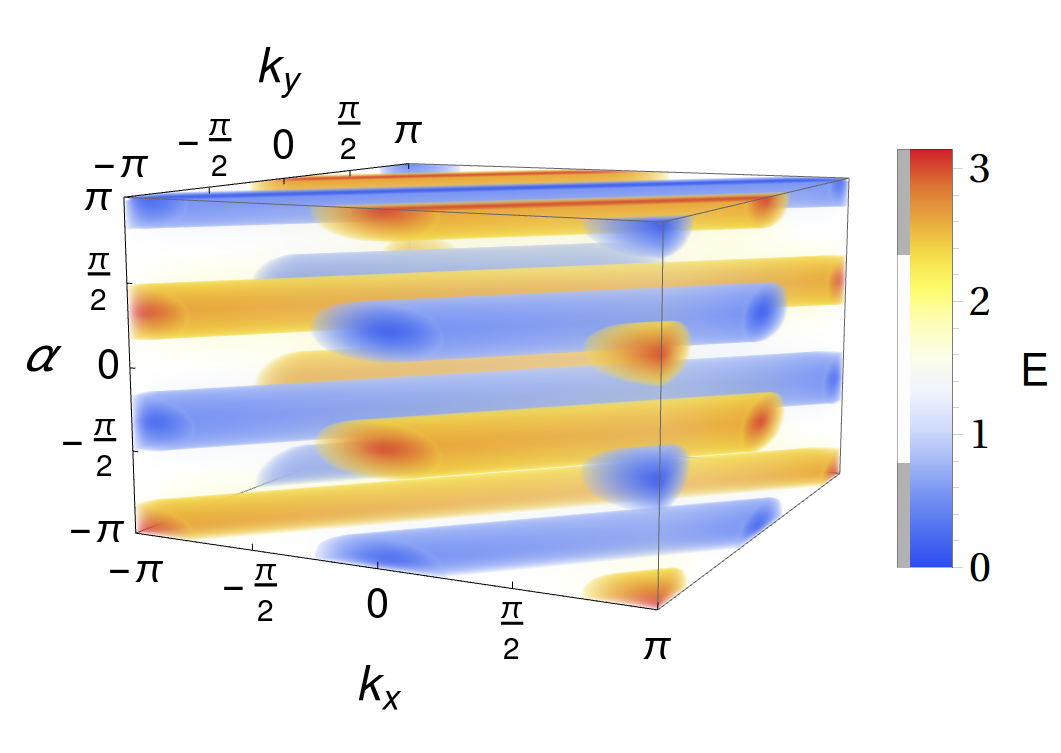}}\sidesubfloat[\label{E-Split-2}]{\includegraphics[width=0.3\linewidth]{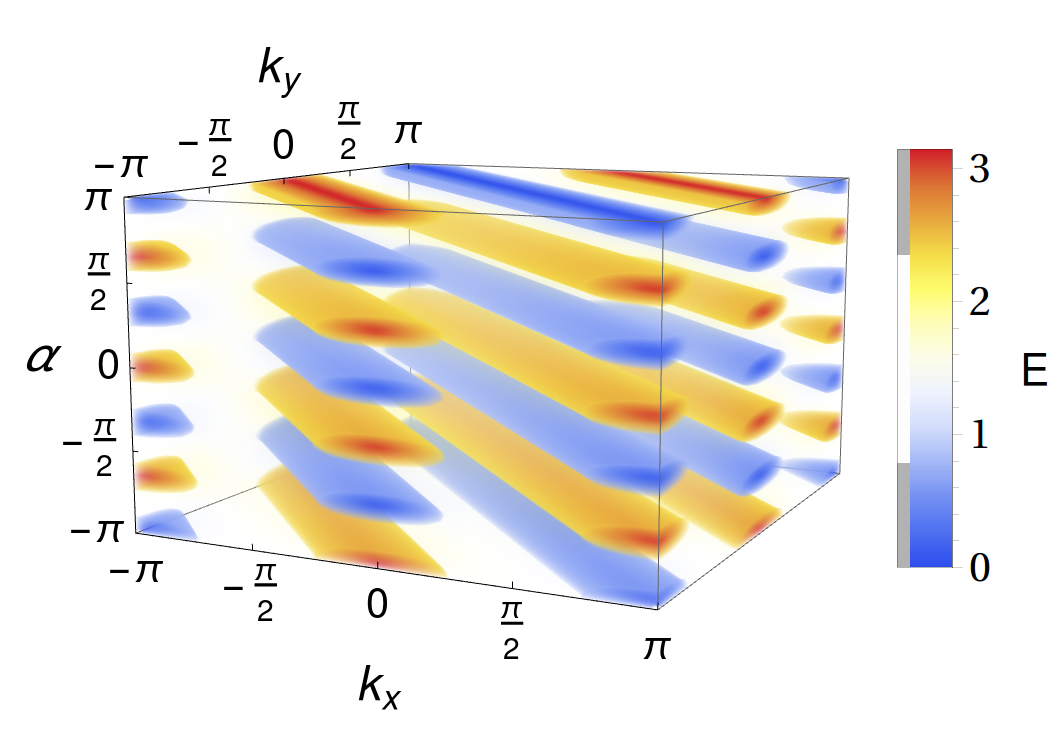}}
			\sidesubfloat[\label{E-Split-3}]{\includegraphics[width=0.3\linewidth]{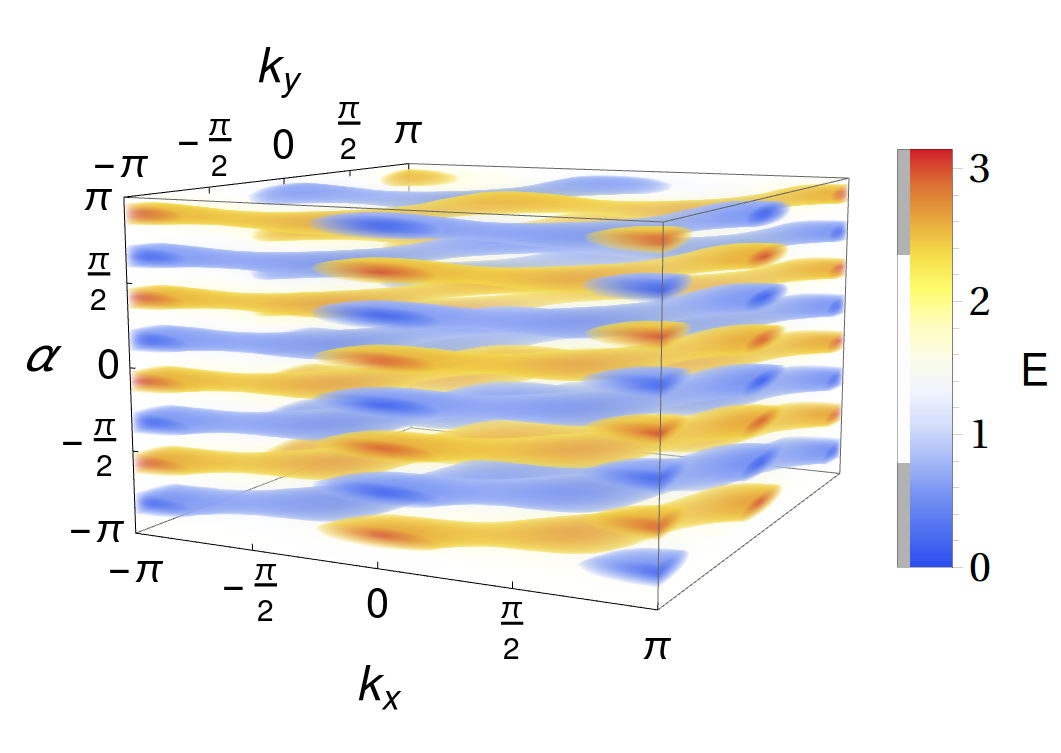}}
			\\[0.0001cm]
			\sidesubfloat[]{\includegraphics[width=1\linewidth]{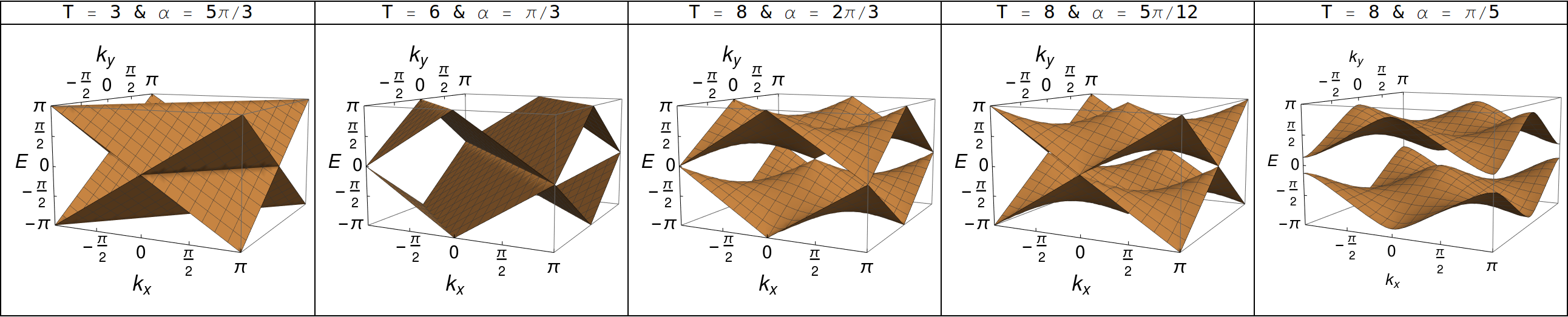}}		
	\end{tabular}}				
	\caption{Split-step quantum walk (two-dimensional): Energy as a function of rotation angle and momenta for $\beta=\pi/3$ with $T=3$ (a), $T=6$ (b) and $T=8$ (c). The simulated topological phenomena for each of these steps has specific characteristic behavior that differ from other steps. In (d), setting $T=3$ and $\alpha=5\pi/3$ results into simulation of Dirac cone and flat bands gapless energy bands. For $T=6$ and $\alpha=\pi/3$, Dirac cones and flat bands are formed. But the flat bands in this case differ from the other case due to different conditions on momenta for having flat bands. By considering $T=8$ with rotation angles $\alpha=2\pi/3$ and $\alpha=5\pi/12$, we observe energy gaps closing linearly (Dirac cone) and nonlinearly (Fermi arc). If $T=8$, $\alpha=\pi/5$, we observe topological phases (gaped energy bands).} \label{Fig6}
\end{figure*}	
\begin{figure*}[htb]
	\centering
	{\begin{tabular}[b]{cc}%
			\sidesubfloat[]{\includegraphics[width=0.3\linewidth]{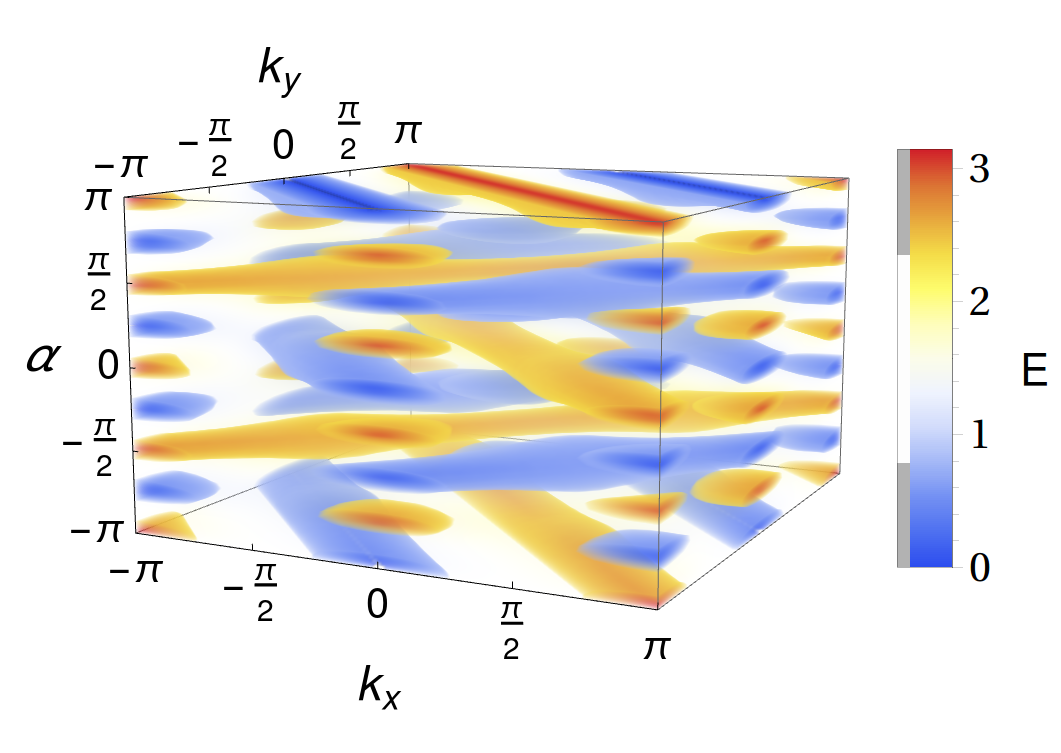}}
			\sidesubfloat[]{\includegraphics[width=0.3\linewidth]{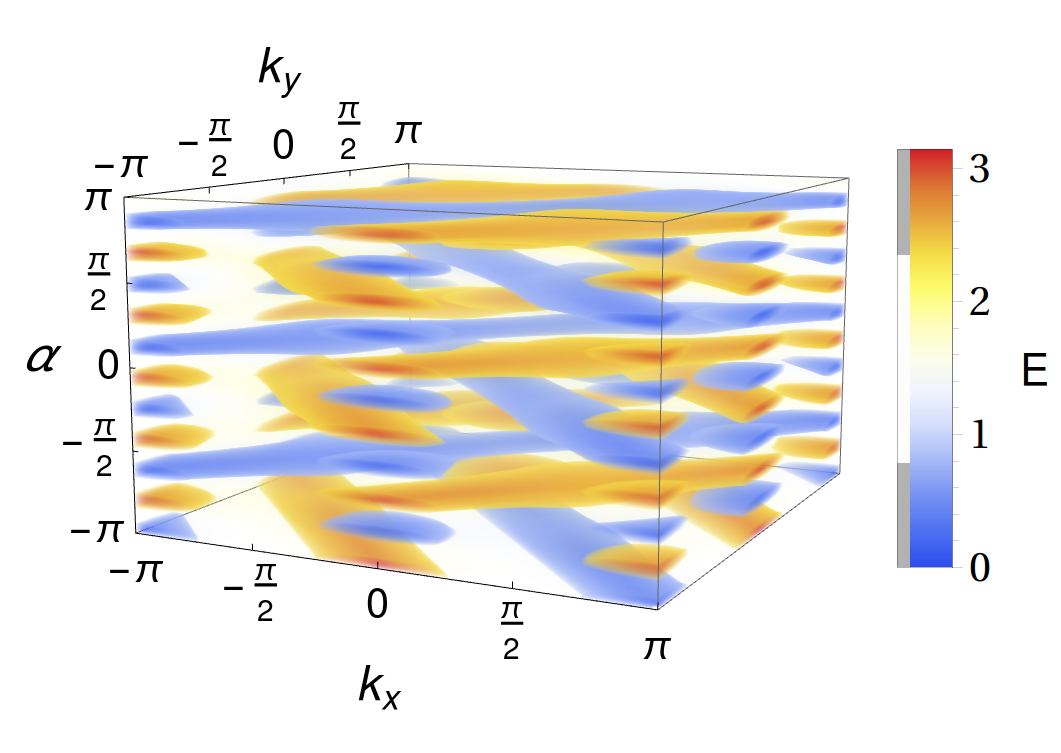}}\\
			\sidesubfloat[]{\includegraphics[width=0.86\linewidth]{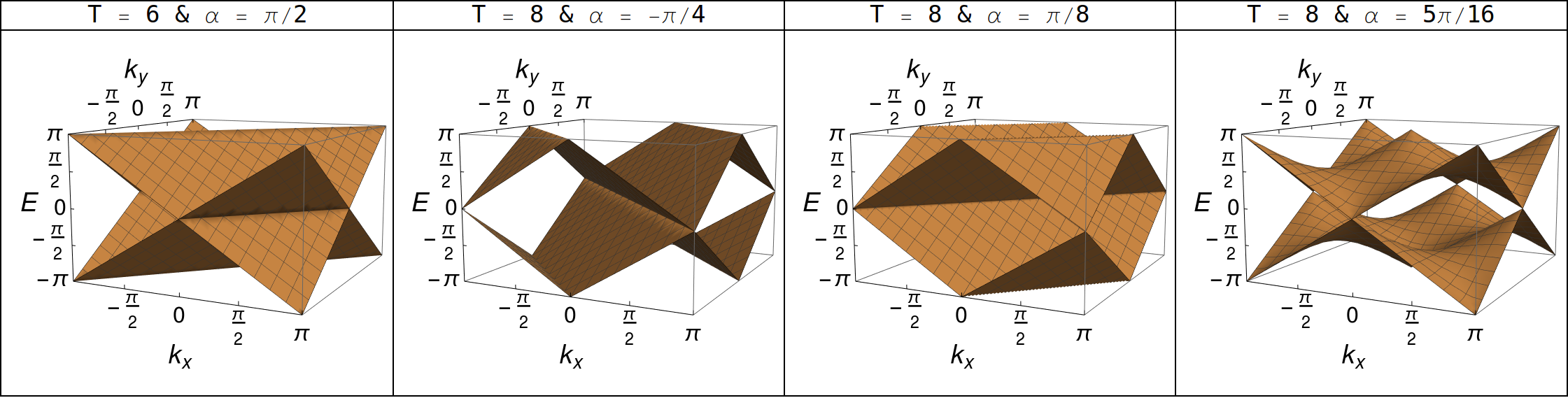}}\\
			\sidesubfloat[]{\includegraphics[width=1\linewidth]{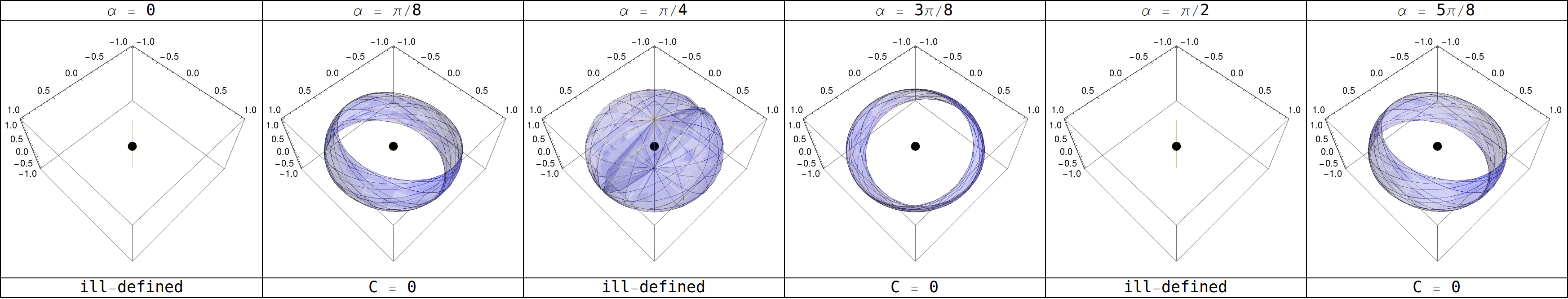}}	
	\end{tabular}}				
	\caption{Split-step quantum walk (two-dimensional): Energy as a function of rotation angle and momenta for $\beta=(\alpha+\pi)/3$, $T=6$ (a) and $T=8$ (b). In (c), for $T=6$ with $\alpha=\pi/2$ or $T=8$ with $\alpha=-\pi/4$ and $\alpha=\pi/8$, gapless energy bands are emerged in forms of Dirac cones (provided $k_{y}=-\pi$) and flat bands (provided $k_{x}$ and $k_{y}$ traverse first Brillioun zone). In contrast, by setting $T=8$ with $\alpha=5\pi/16$, quantum walks simulate Dirac cones and Fermi arcs. Focusing on (a) for $\alpha \in [-\pi/2,\pi/2]$, we notice a cell-like structure for simulated topological phenomena. The cell could be recognized by two flat bands playing the role of the cell's walls, two Fermi arcs and one additional flat bands located between the Fermi arcs. Each cell also contains Dirac cones if one of the momenta is fixed ($k_{y}=-\pi$) and the other one varies. In (d), $\boldsymbol n(k_{x},k_{y})$ is plotted for first Brillouin zone. For gapless energy bands, $\boldsymbol n(k_{x},k_{y})$ passes origin while for gapped phases, it goes around the origin and since it can not cover the origin completely, the Chern number is zero.} \label{Fig7}
\end{figure*}	

\subsection{split-step Quantum walk}  \label{split2}

In this section, we modify the simple-step protocol of the two-dimensional quantum walk into a split-step one and investigate the types of topological phases and boundary states that the split-step quantum walk can simulate. 

There are several types of split-step protocol that we can employ. In this paper, we use the following split-step protocol \cite{Kitagawa,Kitagawa2012,Wang}

\begin{eqnarray}
\widehat{U} & = & \widehat{S}_{\uparrow \downarrow}(y) \widehat{C}_{\alpha} \widehat{S}_{\uparrow \downarrow}(x) \widehat{C}_{\beta}  \label{protocol4},
\end{eqnarray}
where in a single step of the quantum walk, the internal states of walker are rotated by $\widehat{C}_{\beta}$, then it is displaced in $x$ space with $\widehat{S}_{\uparrow \downarrow}(x)$, additional rotation of internal states with $\widehat{C}_{\alpha}$ and finally, displacement of its position in $y$ space with $\widehat{S}_{\uparrow \downarrow}(y)$. 

The coin operators are given in Eqs. \eqref{coin2} and \eqref{coin3}, while the shift operators are $\widehat{S}_{\uparrow \downarrow} (x)= e^{ik_{x} \sigma_{z}}$ and $\widehat{S}_{\uparrow \downarrow} (y)= e^{ik_{y} \sigma_{z}}$. We obtain the eigenvalues of the $\widehat{U}$ in Eq. \eqref{protocol4} as 

\begin{eqnarray}
	\lambda(k)= \gamma \pm \sqrt{\gamma^2-1},
\end{eqnarray}
where $\gamma=\cos(k_{x}+k_{y}) \cos(\frac{T \alpha}{2})\cos(\frac{T \beta}{2})-\cos(k_{x}-k_{y})\sin(\frac{T \alpha}{2})\sin(\frac{T \beta}{2})$ and consequently, we find the energy as  

\begin{eqnarray}
E(k) & = & \pm\cos^{-1}(\gamma), \label{energy4}
\end{eqnarray}
in which $\pm$ represents existence of two bands of energy separated by a gap. The energy traverses $[-\pi,\pi]$ and the energy bands become gapless only at $E=0$ and $E=\pm \pi$. In the next section, we show that such quantum walk can simulate three different types of boundary state. For now, we highlight three interesting cases ($c$ is integer):

I) If $\alpha=\beta=\pm 2 c \pi/T$, then energy bands become linearly dependent on $k_{x}$ and $k_{y}$ given by $E=\pm (k_{x}+k_{y})$.

II) If $\alpha=\beta=\pm (2c+1) \pi/T$, then energy bands again are linear functions of $k_{x}$ and $k_{y}$ given by $E=\pm \cos^{-1}(-\cos(k_{x}-k_{y}))$. 

III) If $\alpha=\pm (2c+1) \pi/T$ and $\beta=\pm 2 c \pi/T$ or vice versa, the energy bands become flat with $E=\pm \pi/2$.

Next, it is a matter of calculation to find $\boldsymbol n(k_{x},k_{y})$ and different components of the group velocity as 
\begin{widetext}
\begin{equation}
n_{x}(k)  =  \frac{\sin (k_{x}+k_{y})\cos (\frac{T\alpha}{2}) \sin (\frac{T\beta}{2})-\sin (k_{x}-k_{y})\sin (\frac{T\alpha}{2}) \cos (\frac{T\beta}{2})}{\sin E(k_{x},k_{y})}, \notag
\end{equation}
\begin{equation}
n_{y}(k)  =   \frac{\cos (k_{x}+k_{y})\cos (\frac{T\alpha}{2}) \sin (\frac{T\beta}{2})+\cos (k_{x}-k_{y})\sin (\frac{T\alpha}{2}) \cos (\frac{T\beta}{2})}{\sin E(k_{x},k_{y})}, \notag
\end{equation}
\begin{equation}
n_{z}(k)  = -  \frac{\sin (k_{x}+k_{y})\cos (\frac{T\alpha}{2}) \cos (\frac{T\beta}{2})-\sin (k_{x}-k_{y})\sin (\frac{T\alpha}{2}) \sin (\frac{T\beta}{2})}{\sin E(k_{x},k_{y})},
\end{equation}

\begin{eqnarray}
V(k_{x})& = & \pm \frac{\sin (k_{x}-k_{y})\sin (\frac{T\alpha}{2}) \sin (\frac{T\beta}{2})-\sin (k_{x}+k_{y})\cos (\frac{T\alpha}{2}) \cos (\frac{T\beta}{2})}{ \sqrt{1- \gamma^2}}, \label{groupv4}
\end{eqnarray} 
\begin{eqnarray}
V(k_{y})& = & \pm \frac{-\sin (k_{x}-k_{y})\sin (\frac{T\alpha}{2}) \sin (\frac{T\beta}{2})-\sin (k_{x}+k_{y})\cos (\frac{T\alpha}{2}) \cos (\frac{T\beta}{2})}{ \sqrt{1- \gamma^2}}. \label{groupv5}
\end{eqnarray}

\end{widetext}

Evidently, the $\boldsymbol n(k_{x},k_{y})$ and group velocity become ill-defined at the gapless points of the energy bands. Therefore, the boundary states could be recognized by ill-defined $\boldsymbol n(k_{x},k_{y})$ (hence ill-defined topological invariant) and group velocity. Finally, we should point it out that Chern number for this protocol is $0$. This can be also shown by plotting $\boldsymbol n(k_{x},k_{y})$ for the first Brillouin zone (see Fig. \ref{Fig7}g).

\subsection{Results} 

Similar to its on-dimensional counterpart, irrespective of simple- or split-step protocol, the Hamiltonian, energy bands, group velocity and $\boldsymbol n(k_{x},k_{y})$ are step-dependent. Therefore, the observed step-dependent nature in case of one-dimensional quantum walks is also present for two-dimensional quantum walk. The presence of the additional momentum provides us with two options regarding type of the gapless energy bands: a) we can fix one of the momenta and let the other one varies through the first Brillouin zone. b) both of the momenta traverse through the first Brillouin zone. We will look at the gapless energy bands for these two specific scenarios.  

For simple-step protocol (see Fig. \ref{Fig5}a), if we fix one of the momenta, the energy bands close their gap linearly, similar to Dirac cones in one-dimensional quantum walk. In contrast, if both of the momenta varies, we observe characteristic flat bands for gapless energy bands. The major differences between these flat bands with their one-dimensional counterpart is that they are momenta dependent and obtained only when $k_{x}+ k_{y}=0$ and $\pm \pi$ (see Figs. \ref{Fig5}b). Therefore, we observe two types of gapless energy bands depending on which approach we take. In addition, due to dependency of energy bands on steps, the places of gapless energy bands and their number change step dependently. This shows that we can use the step number as a mean to engineer the place of topological phases and boundary states.  

Modification to split-step protocol results into an additional term in the obtained energy bands ($\cos (k_{x}-k_{y})\sin(\frac{T \alpha}{2})\sin(\frac{T \beta}{2})$). This additional term plays major role in simulation of boundary states in which the energy bands close their gap nonlinearly, hence simulation of Fermi arc boundary states. Therefore, the emergence of this additional term enables our quantum walk in two-dimensional position space to simulate different boundary states.

In split-step two dimensional quantum walk, if we fix one of the rotation angles and let the other one varies, we observe that in each step, the simulated topological phenomena has specific characteristic behavior that differs from other steps (see Figs. \ref{Fig6}a, b and c). This is the reminisce of similar one-dimensional quantum walk where in each step, only one type of boundary state was simulated (see Fig. \ref{Fig2}a). The major difference is that in the two-dimensional case for each step, we simulate two types of boundary states. The Dirac cones are observed in different steps and it happens if we fix one of the momenta and modifies the other one (see Fig. \ref{Fig6}d). On the other hand, when both of the momenta traverse the first Brillouin zone, the energy bands could close their gap nonlinearly (see Figs. \ref{Fig6}f and g), (Fermi arc boundary states) or flat bands boundary states are observed (see Fig. \ref{Fig6}d).

Inspired by similar one-dimensional split-step quantum walk, we can consider one of the rotation angles to be linearly related to the other one ($\beta= s_{1} \alpha + s_{2}$). The first important consequence of such consideration is simulation of boundary states in three different types of Dirac cone, flat bands and Fermi arcs in a single step (see Figs. \ref{Fig7}a and b). The Dirac cones happens if we fix one of the momenta (see Fig. \ref{Fig7}c) whereas we observe the other two boundary states and flat bands when both of the momenta scan the first Brillouin zone. The second important consequence of such consideration is observation of another type of cell-like structure for simulated topological phenomena (see Fig. \ref{Fig7}a for $T=6$ and $\alpha \in [-\pi/2,\pi/2]$). Each cell is characterized by two flat bands boundary states playing the role of the cell's walls, two Fermi arc boundary states and one additional flat bands boundary state located between the Fermi arc boundary states. It should be noted that the flat bands between the Fermi arcs are different from those of cell's wall. Each cell also contains Dirac cones provided that one of the momenta is fixed and the other one varies. Therefore, in a single cell, we can simulate three different boundary states. Finally, using the step-dependent nature of the energy bands, by the tuning step number, we can determine the number of boundary states or topological phases, their type and where they should be. Therefore, we have high level of controllability over simulation of topological phenomena. 

It is worthwhile to mention that the both of protocols that we used for two-dimensional quantum walk yield trivial phases which have Chern number $0$ (see Fig. \ref{Fig7}d). At the boundary states, the Chern number is ill-defined since $\boldsymbol n(k_{x},k_{y})$ is ill-defined. At the first step, $T=1$, the obtained Hamiltonian, energy bands and other properties results into split-step two-dimensional quantum walk with step-independent coin. In order to simulate topological phases with different Chern number, one should employ other protocols for quantum walk which were introduced by Kitagawa \textit{et al.} in Refs. \cite{Kitagawa,Kitagawa2012}. 

\section{Symmetries}

In this section, we briefly discuss the symmetries that different effective Hamiltonian. We will confirm that these effective Hamiltonians have three symmetries of particle-hole, chiral and time-reversal.

The matrix elements of the protocols that we used for our quantum walks (Eqs. \eqref{protocol1}, \eqref{protocol2}, \eqref{protocol3} and \eqref{protocol4}) are all real. Therefore, the complex conjugate of these protocols are equal to them ($\widehat{U}^{\ast}=\widehat{U}$). Based on the relationship between the protocol of the quantum walk and effective Hamiltonian \eqref{Hamiltonian}, we find 

\begin{eqnarray}
\widehat{H}^{\ast}(k) = -\widehat{H}(k). 
\end{eqnarray}

This indicates that our effective Hamiltonians (irrespective of being simple- or split-step and being one- or two-dimensional) possess particle-hole symmetry, since we can find an antiunitary operator, $\widehat{\mathcal{P}}\equiv \widehat{K}$, in which $\widehat{K}$ is the complex conjugation operator satisfying \cite{Kitagawa,Asboth,Rakovszky} 

\begin{eqnarray}
\widehat{\mathcal{P}}\widehat{H}(k)\widehat{\mathcal{P}}^{-1}= -\widehat{H}(k). 
\end{eqnarray}

The effective Hamiltonians that we simulated by different protocols of the quantum walk have time-reversal symmetry as well. The existence of this symmetry could be confirmed through two methods. First of all, we showed that the obtained energies for different effective Hamiltonians (Eqs. \eqref{energy1}, \eqref{energy2}, \eqref{energy3} and \eqref{energy4}) have a property in which $E(k)=E(-k)$. This is one of the requirements for the presence of time-reversal symmetry \cite{Kitagawa}. In addition, we confirmed that the protocols of the one-dimensional quantum walks have chiral symmetry. It is straightforward to prove the presence of such symmetry for two-dimensional cases as well. The presence of particle-hole and chiral symmetries guarantees that there is an antiunitary operator, $\widehat{\mathcal{T}}$, which satisfies \cite{Kitagawa} 

\begin{eqnarray}
\widehat{\mathcal{T}}\widehat{H}(k)\widehat{\mathcal{T}}^{-1}= \widehat{H}(k), 
\end{eqnarray}
where $\widehat{\mathcal{T}}\equiv \widehat{\Gamma}\widehat{\mathcal{P}}$. 

The presence of these symmetries in these forms indicates that one-dimensional quantum walk  with protocols \eqref{protocol1} and \eqref{protocol2} can simulate all type of the topological phases presented in BDI family of topological phases. For two-dimensional case, the simulable phases are trivial ones with Chern number $0$. This is due to symmetries present in system.

Finally, it is worthwhile to take a look at the chiral symmetry operator of one-dimensional split-step quantum walk \eqref{chiralsplit} at different types of boundary (or edge) state when $\beta=(\alpha+\pi)/3$. In case of Dirac cone boundary (or edge) states, for the rotation angles where energy bands close their gap, the chiral symmetry operator reduces to $\widehat{\Gamma}= (-1)^{T/2}\sigma_{x}$. As for the flat band boundary (or edge) states, we obtain the chiral symmetry as $\widehat{\Gamma}= -\sigma_{z}$. Additionally, if the boundary (or edge) states are Fermi arcs, the chiral symmetry operator yield $\widehat{\Gamma}=\frac{1}{\sqrt{2}} [(-1)^{T/2}\sigma_{x}-\sigma_{z}]$. We observe that for each type of boundary (or edge) states, we have specific characteristics that enable us to distinguish them from one another and put each boundary (or edge) state in its proper category. 

\section{Further possibilities}

As a final remark, we return to the flexibility in the split-step protocol. The split-step protocol considered in this paper is a straightforward generalization from simple-step quantum walk to split-step one. We can consider the situations where the first or the second coin operator is step-independent in split-step protocol. This significantly will change the Hamiltonian, energy and other properties of the system, hence simulated topological phenomena by the quantum walk. 

A simple example is if we use a step-independent coin for the first coin operator in Eq. \eqref{protocol2}. Previously, we showed that the chiral symmetry operator and $\boldsymbol A$ are determined only by the first coin operator in the protocol. Accordingly, these two quantities would be step-independent while the other properties (Hamiltonian, energy and etc.) would remain step-dependent. 

In addition, we have the possibility to include position-dependency in the protocol of the walk \cite{Ramasesh}. Another route to expand the present work is to employ step-dependent coins for the set of protocols that were introduced and used in Ref. \cite{Cardano}. Each of these protocols are employed to simulate specific family of the topological phases. We can also use the step-dependent coins to build up three-dimensional protocols as well. These are done in Ref. \cite{Panahiyan2020}. 

To detect presence of the boundary states, one can use the earlier work of Cardano et al. \cite{Cardano2017} in which they showed that average chiral displacement of a particle’s wave packet becomes quantized and proportional to the winding number. In addition, in another work \cite{Cardano}, they also showed that statistical moments of the unitary quantum-walk dynamics can also be used to characterize topological phase transitions. Both of these methods can be applied to recognize the presence of the boundary states that we studied in this paper as well as topological phases. Additionally, in a recent work \cite{Panahiyan2020-3}, we showed that critical phenomena simulated in quantum walk (boundary states) can be characterized by critical exponents, length scale and correlation function. In investigation of the correlation function, we used Discrete Fourier Transformation between $k$ space and Wannier states. We showed that correlation function decays via a damped oscillation for Dirac cone boundary states while for Fermi arc type, the decay happens monotonically. This also provide tools for detecting different types of boundary state.

\section{Conclusion}  \label{Conclusion}

In this paper, we used the step-dependent coin in the protocols of the one- and two-dimensional quantum walk to simulate topological phases, boundary and edge states. We considered two types of protocol for the quantum walks known as simple-step and split-step protocols. 

The one-dimensional quantum walks with simple-step protocol simulated only boundary states in form of Dirac cone and two non-trivial topological phases. The step-dependent coin provided the number of steps as a factor which changes the properties of the simulated topological phases (Hamiltonian, energy and etc.). In fact, by tuning the number of steps, we can engineer the type of topological phases, the size of each topological phase and the place of boundary states, hence phase transitions. Therefore, with simple-step quantum walk, we have highly controllable simulation of non-trivial topological phases and Dirac cone boundary states.  

The modification of the simple-step protocol to split-step one significantly enriched the capability of the quantum walk to simulate topological phases, boundary and edge states. First of all, the one-dimensional quantum walk with split-step protocol simulated all types of the boundary and edge states including Dirac cone, Fermi arc and flat bands. In addition, we showed that this quantum walk also simulates all types of topological phases in BDI family. Therefore, the quantum walk with split-step protocol and step-dependent coin can simulate all types of BDI family of the topological phases, boundary and edge states (and corresponding topological phase transitions). 

In generalization to two-dimensional quantum walk with simple-step protocol, we simulated two types of boundary state at each step: Dirac cones if we fix one of the momenta and let the other one varies through the first Brillouin zone, and momenta-dependent flat bands boundary states if both of the momenta traverse the first Brillouin zone. In contrast, we showed that if simple-step protocol is modified to split-step in two-dimensional quantum walk, we can simulate all three types of Dirac cone, flat bands and Fermi arc boundary states.   

Finally, we showed that the rotation angles of coin operators and number of steps are controlling factors. They can be used to enforce simulation of only one (two) type(s) of the boundary state at each step or all three types of the boundary states together at specific steps. Moreover, we were able to simulate exotic cell-like structures for the topological phases and boundary states. Each cell contained all types of boundary states (and corresponding topological phase transitions). The step-dependent coins enabled us to determine the place of boundary states or topological phases, their types and their numbers. This provides us with highly controllable simulation of topological phenomena and allows one to achieve the universal simulator of topological phases and boundary states with quantum walk.

\end{document}